\documentclass[a4paper]{article}

\usepackage[english]{babel}
\usepackage[utf8x]{inputenc}
\usepackage[T1]{fontenc}

\newcommand{\be}{\begin{equation}} 
\newcommand{\ee}{\end{equation}}  
\newcommand{\bea}{\begin{eqnarray}}
\newcommand{\eea}{\end{eqnarray}}

\newcommand{\f}[2]{\frac{#1}{#2}}
\newcommand{\bup}[1]{\left(#1\right)}
\newcommand{\rup}[1]{\left[#1\right]}

\usepackage[a4paper,top=2.5cm,bottom=2.5cm,left=2.5cm,right=2.5cm,marginparwidth=1.75cm]{geometry}

\usepackage{chngpage} % TO CHANGE  margins to a specific table that was going outside them
\usepackage{booktabs}
\usepackage{amsmath}
\usepackage{graphicx}
\usepackage[colorinlistoftodos]{todonotes}
\usepackage[colorlinks=true, allcolors=blue]{hyperref}
\usepackage{topcapt}
\usepackage{amssymb}
\usepackage{natbib}
\usepackage{url,babel}
\usepackage{textcase}
\usepackage{graphicx}
\usepackage{amsfonts,amsmath}

\title{Are `Water Smart Landscapes' Contagious? An epidemic approach on networks to study peer effects.\footnote{This manuscript has been co-authored by UT-Battelle, LLC, under contract DE-AC05-00OR22725 with the US Department of Energy (DOE). The US government retains and the publisher, by accepting the article for publication, acknowledges that the US government retains a nonexclusive, paid-up, irrevocable, worldwide license to publish or reproduce the published form of this manuscript, or allow others to do so, for US government purposes. DOE will provide public access to these results of federally sponsored research in accordance with the DOE Public Access Plan (\url{http://energy.gov/downloads/doe-public-access-plan}).}}

\author{Christa Brelsford$^{1}$, Caterina De Bacco$^2$ \\ \href{brelsfordcm@ornl.gov}{brelsfordcm@ornl.gov}, \href{mailto:cdebacco@santafe.edu}{cdebacco@santafe.edu} }
\date{$^1$Oak Ridge National Laboratory, 1 Bethel Valley Road, Oak Ridge, TN 37830  \\ $^2$Santa Fe Institute, 1399 Hyde Park Road, Santa Fe, NM 87501, USA}

\begin{document}
\maketitle

\begin{abstract}

We test the existence of a neighborhood based peer effect around participation in an incentive based conservation program called `Water Smart Landscapes' (WSL) in the city of Las Vegas, Nevada. We use 15 years of geo-coded daily records of WSL program applications and approvals compiled by the Southern Nevada Water Authority and Clark County Tax Assessors rolls for home characteristics.  We use this data to test whether a spatially mediated peer effect can be observed in WSL participation likelihood at the household level. We show that epidemic spreading models provide more flexibility in modeling assumptions, and also provide one mechanism for addressing problems associated with correlated unobservables than hazards models which can also be applied to  address the same questions.  We build networks of neighborhood based peers for 16 randomly selected neighborhoods in Las Vegas and test for the existence of a peer based influence on WSL participation by using a Susceptible-Exposed-Infected-Recovered epidemic spreading model (SEIR), in which a home can become infected via autoinfection or through contagion from its infected neighbors. We show that this type of epidemic model can be directly recast to an additive-multiplicative hazard model, but not to purely multiplicative one. Using both inference and prediction approaches we find evidence of peer effects in several Las Vegas neighborhoods.

%Using a dataset of 100,000 households in Las Vegas, Nevada, we show that one household's participation in the `Water Smart Landscapes' program is likely to have had some influence on their neighbors' probability of participation.  We use an SI epidemic model as well as an additive multiplicative hazard model explore the role of social contagion in program uptake.  

%The main analytical methods are from network epidemiology, but we also compare the results from traditional economic hazard models with approaches from network epidemiology. We show how hazard models make some implicit assumptions about transmission mechanisms through networks that may not always be true.   

\end{abstract}

\clearpage

\section{Introduction}
In the face of increasing climate variability and growing water demand associated with rising populations, policymakers are faced with the harsh reality of water scarcity. Conservation measures targeting outdoor landscaping have become popular because consumers often are unaware of their outdoor water use, suggesting that substantial savings may be generated with even small incentives and changes in customer awareness. ``Cash for grass'' style water conservation programs have become an important water conservation strategy in the western US over the last two decades. 
The Southern Nevada Water Authority's (SNWA) Water Smart Landscapes program (WSL) pays homeowners to replace their lawns with xeric (desert) landscapes, one of the longest running ``cash for grass'' programs. 
While the difference in watering requirements of mesic vs. xeric landscaping are well established, and short-run savings have been demonstrated in a few cases a number of questions are unanswered about turf-removal subsidy programs \cite{brelsford2017growing}.

WSL participation is highly visible to neighboring homes, and may trigger a mechanism of social learning that can impact the neighbors' adoption behavior. 
This stimulates a key research question:  can we observe neighborhood based peer effects in household participation in Las Vegas' Water Smart Landscapes program? 
Measuring the influence that an individual has on their peers is important for estimating the indirect benefits a conservation policy may create, but distinguishing true peer effects from homophily, correlated unobservables, and reflection is econometrically challenging. 
Modeling peer influence on networks through epidemic modeling techniques may provide a useful tool for identifying peer effects without the bias that these challenges can introduce. We analyze a very high resolution dataset from the WSL program in Las Vegas, Nevada.
This dataset includes both the day of application, and the day of completion of all program requirements for each participating household in the Las Vegas Valley Water District Service area. 
These data thus allow us to differentiate between the decision to participate in WSL and its implementation.

The challenges to unbiased identification of peer effects are substantial in the absence of experimental data, and have been well described by \cite{manski1993identification, manski_economic_2000, brock2001interactions, soetevent2006empirics, aral_distinguishing_2009, shalizi_homophily_2011, angrist_perils_2014, ryan_measurement_2017} and others.
There are three major challenges to the econometric identification of peer effects. 
First, reflection between two participating homes or individuals with similar adoption times, makes it difficult to determine which participant influenced the other. 
Second, homophily, where households that are predisposed to participate in the WSL program self-select into the same neighborhoods.  
%This could be because people who share values around water conservation may all select similar neighborhoods. 
Finally, correlated unobservables may also spuriously generate the appearance of peer effects, when in fact, there are spatially mediated unobserved factors which differentially influence households to participate in WSL. 
For example, perhaps local environmental conditions make grass particularly difficult to keep alive due to locally higher temperatures, excessive wind, or poor soil quality.

Several recent and working papers by \cite{bollinger_peer_2012,graziano_spatial_2015,baranzini_what_2017} have explored the role of peer effects in solar panel adoption and have made important progress in developing techniques to address the challenges described by previous authors.
\cite{towe_contagion_2013} identified peer effects among homes in a neighborhood in the context of home foreclosures.  There are important similarities between WSL program adoption and solar panel adoption, but the existence of a peer effect has not been explored in ``cash for grass'' style subsidy programs. 

\cite{bollinger_peer_2012} exploit the lag between the decision to adopt and actual installation of solar panels to address the reflection problem.
\cite{graziano_spatial_2015} handle the definition of peers in a spatially disaggregated manner, which avoids some kinds of boundary problems that aggregating to a geopolitical boundary induced in \cite{bollinger_peer_2012}.  \citeauthor{graziano_spatial_2015} also use a rich set of spatial and temporal fixed effects to address correlated unobservables. 
The core analytical method used by each of \citeauthor{bollinger_peer_2012,graziano_spatial_2015} and in the \cite{baranzini_what_2017} working paper are variations on linear regression with fixed effects at an appropriate level of spatial aggregation.
They estimate the space and time specific fraction or count of eligible places that may adopt solar panels in that time step.  
This method does not explicitly address the fact that solar panel adoption is a one-time event, and places that have already installed solar panels cannot do so again, no matter how much stronger the influential factors may become. 
Thus, hazard models such as that employed by \citeauthor{towe_contagion_2013} may be conceptually more appropriate to study peer effects in one-time behaviors. 

The Cox proportional hazard model \citeauthor{towe_contagion_2013} employ implicitly requires a strictly multiplicative relationship between the various factors influencing the household hazard rate.
An additive or additive-multiplicative relationship between the various factors that influence participation (as allowed in the linear models employed by Gillingham and co-authors) is desirable. A purely multiplicative model forces the hazard rate to zero if either one of the epidemic or endemic effects is zero. This does not happen in models where the hazard rate is instead a superposition of epidemic and endemic effects, as in our case or additive-multiplicative survival models as in  \cite{hohle2005inference,hohle2008spatio,hohle2009additive}. An additive-multiplicative model, which allows for non-zero probabilities of participation when some factors are zero might be more appropriate in cases like ours, where we want to allow a baseline participation probability to contribute to the hazard rate even in the event the peer effect is zero (and vice versa).  However, the types of hazards models that have been employed in the economics literature cannot be recast into an additive-multiplicative model, while SEIR style models from mathematical epidemiology can be. 

In order to combine and extend the novel contributions of each of these papers, we follow \citeauthor{bollinger_peer_2012}'s approach to address reflection issues by using the gap between the decision to participated in the WSL program and actual adoption.
We fully exploit the spatially rich parcel level nature of our dataset to develop an accurate, household specific model of peers, additionally exploring the importance of both Euclidean and on-road travel distance to define the set of peers attributed to each home. 
Finally, drawing from and extending the ideas used by \citeauthor{towe_contagion_2013} to select a hazard model as the main choice for inference, we model the influence of peers using an epidemic modeling approach on networks at the household level which permits an additive/multiplicative model of influence. 

The language of contagion, infection, and transmission has been widely used in economic research on peer effects but the application of the mathematical tools developed for the purpose of modeling disease transmission in this context is novel.
The epidemic modeling approach we use allows the use of an additive/multiplicative model of influence instead of a strictly multiplicative model, even within a hazards context. 
It also permits complete flexibility in modeling transmission dynamics on networks and makes the implicit network structure used in the aggregate linear models completely explicit. 
This allows adoption patterns on the actual network to be simulated, so predicted transmission dynamics can be compared to actual dynamics.  
Because self-sorting into neighborhoods (homophily) and other correlated unobservables change slowly relative to the rate of program adoption, comparison of simulated temporal dynamics with or without a peer effect to actual temporal dynamics should provide evidence about the existence of a peer effect without bias from homophily and correlated unobservables. 
Finally, networked epidemic models of peer effects open an opportunity to explore a variety of factors that have been considered in epidemiological studies, but not in a conservation program adoption context. For example, this method could be used to study how spatial properties of neighborhoods and cities might influence program adoption; or allow simulation for how targeted campaigns may speed program adoption. 

In this paper, we strive to use the language of both epidemiology and economics in a way that is intelligible to readers in either field.  A home that has had a landscape conversion recorded through WSL can be referred to as \textit{active} or \textit{infected}.  In this paper, the \textit{epidemic} effect is synonymous with the economic term \textit{peer effect}:  the role that one home's WSL adoption behavior plays in influencing their neighbors' eventual participation probabilities.  The \textit{endemic} effect contains all of the reasons a household might participate in WSL except for the influence of their neighbors' participation, and is thus a major component of an individual homes activation probability.  A home that \textit{autoinfects} is one that becomes active independent from the influence of their neighbors. 

We adapt epidemic modeling tools to the inference of peer effects by mathematically modelling the dynamics of participation in the WSL program as a discrete-time epidemic spreading model SEIR with \textit{autoinfection}. 
In contrast with standard epidemic models, we allow a house to activate, i.e. adopt the program, because of a combination of \textit{endemic} causes in addition to the standard \textit{epidemic} transmission, where infection is `transmitted' along the network by a neighboring house through the process of social influence.

A \textit{susceptible} home in this model is one that has no recorded prior WSL conversions. An \textit{exposed} home is one that has submitted an application for a WSL conversion to the SNWA, and thus has already made the decision to activate, but has not yet completed the landscape conversion process.  \textit{Infectious} homes have completed the landscape conversion process as required by SNWA, and so their changed landscape is visible to their neighbors. A home that has \textit{recovered} is one in which the WSL conversion has happened sufficiently long ago that it no longer influences its neighbors activation probability, not that the home has reverted from a xeric landscape back to turf.

We find evidence of peer effects in several of the sample neighborhoods. The presence of WSL participating neighbors increases a non-participating house's probability of adoption. 
For neighborhoods where we see a peer effect, the inclusion of this effect in addition to the endemic effect means that we both obtain better model likelihoods, and we are also better able to predict the evolution of adoption dynamics for several months after the observation period. These two strategies for identifying the best model will not necessarily provide the same result.
In addition, we find that a WSL participating neighbors' influence on their peers fades in time with finite recovery rates on the order of a few months to a year, after which the influence fades.

The remainder of the paper is structured as follows. Section 2 covers the relevant history of the WSL program and other important characteristics of the adoption environment in Las Vegas during the study period.  Section 3 describes the dataset we use.  Section 4 describes the analytical methods used.  Section 5 presents the results of the conditional probability estimates, as well as inference and prediction performance results.  Section 6 concludes.

\section{Program Background \& Context}

Las Vegas relies heavily on the Colorado River for its urban water supply.  
About 90\% of Las Vegas' water supply is withdrawn from Lake Mead, created in 1936 with the construction of the Hoover Dam.
Withdrawals are strictly regulated through the Colorado River Compact and other related interstate agreements on how Colorado River water is shared. 
The Colorado River Compact allots Nevada a fixed annual withdrawal of 370 million m$^3$ (300,000 acre-feet) of water from the Colorado River.  
Las Vegas' treated waste water is returned to lake Mead via the Las Vegas Wash, and the Metro area is given full credit for all water that is returned back to Lake Mead. 
Nearly all water that is used indoors in Las Vegas flows through the sewer system and is returned to lake Mead, while nearly all water that is lost to evapo-transpiration.  
This return flow credit creates a strong incentive for Las Vegas to focus their conservation policy on reducing outdoor water consumption, and WSL has been the cornerstone of Las Vegas' outdoor water conservation strategy.  

\begin{figure}[htb]
   \centering
   \includegraphics[width=4in]{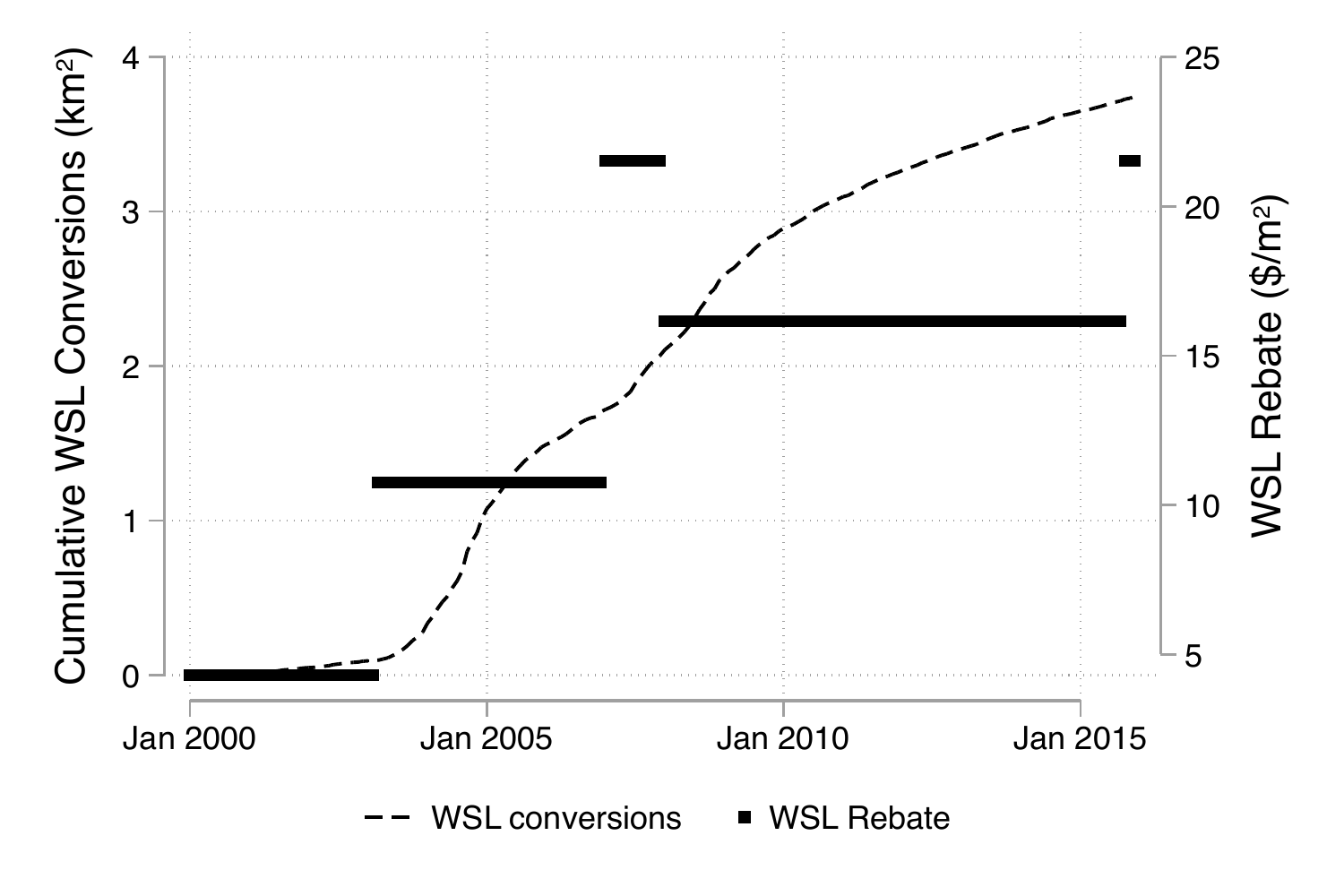}
   \caption{Cumulative WSL conversion area in acres and the nominal WSL rebate at that time. We group WSL participants into four cohort groups based on the nominal rebate price they received. Cohort 1 includes households that participated before February 2003, Cohort 2 includes households that participated between March 2003 and December 2006, Cohort 3 includes January to December 2007 conversions, and Cohort 4 includes households that participated after January 2008. }
   \label{fig:WSL_summary}
\end{figure}

Southern Nevada Water Authority (SNWA) began the WSL program in 1996 as a small pilot program then called Southern Nevada Xeriscapes. 
WSL has always been a completely voluntary, incentive based program.  
Owners receive a rebate check in response for removing grass or swimming pools and replacing it with 'Water Smart Landscapes'. 
Over the course of its history, WSL has paid residential and commercial landowners between \$4.30 and \$21.50 per square meter of grass removed and replaced with xeric landscaping.  
In July 2000 the program took on its modern form by issuing rebates to customers who converted their lawns to desert landscaping based upon the size of the converted area, although it became a widespread and important aspect of Las Vegas' water supply security plan during the 2002-2004 drought. 
Fig.~\ref{fig:WSL_summary} shows the cumulative area of WSL conversions over the program's history, demonstrating how it grew from a relatively small-scale program to a widespread and important aspect of SNWA's water supply security plan after the 2004 drought declaration.
During the programs pilot phase, rebate values were \$4.31/m$^2$ (\$0.40/ft$^2$).  
In February 2003, this was increased to \$10.73/m$^2$ (\$1.00/ft$^2$).  
In December 2006, this was further increased to \$21.51/m$^2$ (\$2.00/ft$^2$).  
In January 2008, the rebate value was reduced to \$16.15/m$^2$  \$1.50/ft$^2$.
This remained until October 2015, when the value was again increased to \$21.51/m$^2$ (\$2.00/ft$^2$).
Throughout the program's history, there have been limits on the maximum rebate available for residential consumers, and a tired structure where the first area converted receives the full rebate value, and additional areas beyond that receive a lower rebate per-square foot.
The tiers and caps on rebate area and total rebate amount are set at high enough values that they are likely to have had little influence on conversion behavior for single family residential homes.\footnote{In August 2000, the cap was 230 m$^2$, which was increased to 5,800 m$^2$ in Feb 2002, and further raised to 46,000 m$^2$, while the median lot size is 650 m$^2$ and the 99th percentile lot is 2,990 m$^2$. Even the earliest cap, in August 2000 still allowed the full rebate to be afforded to 90\% of conversion participants, and per household conversion size has generally fallen, rather than increased, as the program ages. Thus, is appears likely that these caps were primarily targeted at large commercial facilities such as golf courses.}

SNWA notes that typical landscape conversions cost about \$15 per m$^2$ (\$1.40 per square foot in 2000 dollars), although higher end landscapes can cost substantially more \cite{sovocool2006depth,SNWA_FAQ_2018}
This means that for more recent WSL cohorts, the rebate incentive could cover most of the cost of a typical conversion.
A more detailed history of the WSL rebate structure and limits is outlined in \cite{brelsford2014whiskey}. 

SNWA's efforts at marketing the program have also evolved over the programs lifespan.
In its early phase, WSL was marketed at individual community events, and the communication message was generally aimed at describing what Xeriscaping is, and informing residents that the WSL program existed.  
Later, messaging changed to describing what purpose grass serves in a desert environment, communicating with ideas like ``if you only walk on your grass to push a lawnmower, you may want to think about replacing with Xeriscaping'' or ``you do not need wall-to-wall carpeting if an area rug will do!''  
These messages were explained using mass communication and paid advertising.  
More recently, knowledge of both what Xeriscaping, or Water Smart Landscapes are is widespread, and SNWA's WSL focused marketing efforts have been more based on individually targeted mailers concurrently with a semi-annual citywide advertising push to remind consumers to update their irrigation clock with the changing seasons. 
There were substantial targeted mailings sent out in spring and summer 2007, roughly concurrent with the brief period where the WSL rebate was at its highest. 

During the study period, municipal governance in the Las Vegas metro area also passed a number of restrictions on the use and installation of water intensive outdoor features including grass, swimming pools, and other water features.  
In November 2000 the first limits on turf installation in new construction were passed, requiring that front yards be no more than 50\% turf.  
In June 2003, an ordinance was passed that restricts the size of pools and water features permitted in new construction, and also restricts the use of grass to less than half of the back yard area and prohibits it entirely in front yards.  
There have been minor changes in building code since 2003, but the general restrictions as applied to single family residential homes have remained consistent.
Changes to municipal code apply to new construction only. 

 The 2008 recession had a dramatic effect on Las Vegas' economy and home values. 
 Regional GDP fell by 13\% between 2007 and 2010, and recovered to the pre-recession levels in 2015 \cite{FRED_GDP_2018}. 
 Unemployment rose from  4\% to 13\% between 2007 and 2010, and has now declined back to about 5\% \cite{FRED_unemployment_2018}. 
 Between 2006 and 2012, home values fell from 2.5 times the 1995 housing price index back to their 1995 value and have since been slowly recovering \cite{FRED_HPI_2018}. 
 Foreclosure rates rose to nearly 10\% (among the highest in the nation) when 32,000 homes were foreclosed upon in 2009 \cite{hogan_strip_2016}. 
 This also created large numbers of vacant homes, a problem that still challenges the region \cite{segall_number_2017}. By the end of the study period in 2015, the economy had largely rebounded. 
 The recession was associated both with reduced WSL rebate values in Jan 2008, and also a significant slow down in housing stock growth rates, from 16,000 new homes constructed per year at the 2004-2005 peak, to only 2,500 in 2009 \cite{FRED_start_2018}. 
This recession is likely to have influenced WSL participation, but the directionality is unclear. 
Cash-strapped homeowners may have sought out the WSL program for the extra income it could provide, while, once a home is in foreclosure or vacant, it cannot participate. 

\begin{figure}[h]
\centering
\includegraphics[width=12cm]{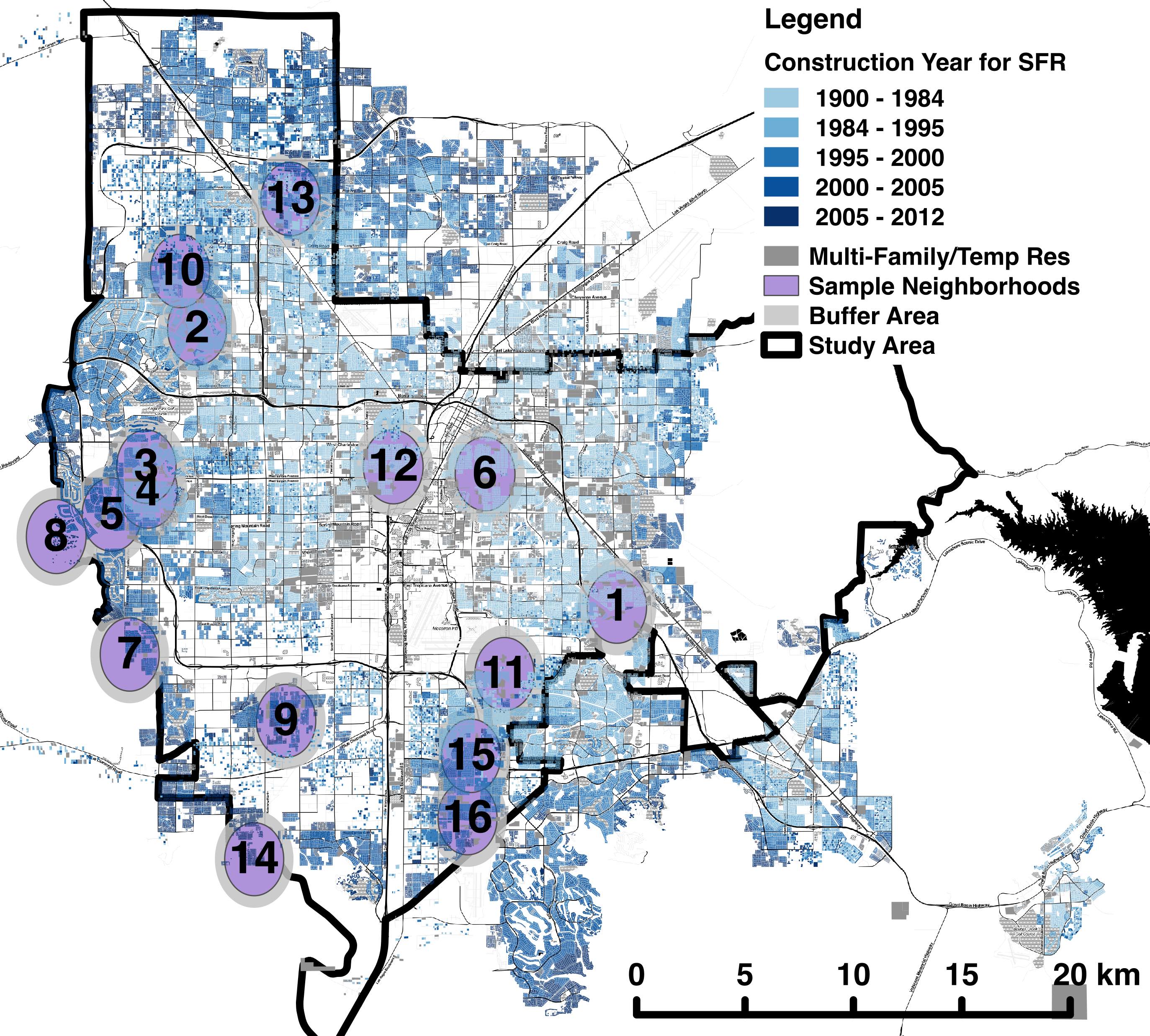}
\caption{Single Family residential parcels, colored by construction year.  Multi-Family or temporary residential structures such as duplexes, apartments, townhouses, condos and mobile homes are colored grey, and not included in this analysis.  Sample neighborhoods are defined in purple, with a light grey buffer zone delineating parcels that are potentially included in descriptions of a focal homes neighbors, but are not themselves included in the analysis.}
\label{fig:map}
\end{figure}

Finally, Las Vegas' rapid population growth and the associated new construction had a substantial influence on the installed housing stock. 
In our dataset, the count of homes grew from 183,000 to 292,000 between 2000 and 2015, with the bulk of the new construction occurring before the 2008 recession. 
Fig. \ref{fig:map} shows all single family residential parcels in Las Vegas, colored by construction year, and demonstrates the substantial growth in residential housing that has occurred in this period.  
The recent growth means that about 25\% of Las Vegas' housing stock was constructed after the 2003 restrictions on turf grass in new construction had been implemented, making them much less attractive candidates for WSL participation because of the initial limits on turf.

\section{Data}
The dataset used in this analysis is a nearly complete cross-section of single family homes in the Las Vegas Valley Water District Service area.  
Fig. \ref{fig:map} shows a map of all households included in the consumption dataset colored by their construction year. 
Out of the 463,658 single family homes in the Clark County Tax Assessors records and 46,079 households in the WSL conservation program records, 299,158 homes (including 29,752 WSL participants) are in the study area and so have matched records of residential WSL participation and home physical characteristics.  
The 16,327 WSL participating households that are not in the study area have similar physical characteristics to the participating homes that are in the study area. 
Each record includes participation behavior and the home's structural characteristics as defined by the Clark County Assessors office in 2012. 
Structural characteristics include indoor area, lot size, number of rooms, bathrooms, bedrooms, and plumbing fixtures, as well as the presence or absence of a pool and 2012 assessed value.  
We exclude value data for 1,867 homes constructed in 2011 and 2012 because the homes did not yet have a valid assessed home value provided by the Assessors Office.
We exclude 36 households because the recorded indoor characteristics for the home are physically impossible. 
Finally, we exclude 3,145 WSL participating households in the study area because they have multiple recorded WSL conversions during the study period; we focus on homes with single conversions for the sake of clean identification.
This provides a complete, cleaned cross sectional dataset of 291,737 homes and 24,206 WSL participants. 

Clark County Assessors office files are used to provide spatial information on both the location of individual homes and also the road networks necessary to calculate on-road travel distances between pairs of homes.

\subsection{Descriptive Statistics}

There is significant spatial and temporal clustering in the physical characteristics of homes. 
%We do not have household level demographic information, but demographic homophily is widely observed and easily explained \cite{schelling_dynamic_1971} and is a significant challenge to the true identification of social contagion.  
There are significant differences in the characteristics of homes which did participate in the WSL program, and those which did not.  WSL participating homes have substantially larger lots, many more pools, are somewhat more valuable, and have small increases in indoor size metrics including indoor area and number of rooms compared to non-participating homes. Participating homes are also located in block groups with higher homeownership rates, which is consistent with a policy incentive structure that primarily rewards homeowners.\footnote{Ownership rates are inferred by attributing to each home the block group level probability of a home being owner occupied vs renter occupied. This method is likely to underestimate the true difference.}  Tab.~\ref{tab:descriptive stats} shows several key characteristics of WSL participating homes and non participating homes for Las Vegas as a whole.  %Results for the sampled neighborhoods are not significantly different, and the only characteristic with a difference greater than 5\% is lot area.  For the sample population, lot area is 599 m$^2$ and 742 m$^2$ for non-participants and participants, respectively. 

\begin{table}[htb]
   \centering
   \topcaption{Descriptive statistics for Las Vegas housing.}
\begin{tabular}{lrrrr} \hline
 & \multicolumn{2}{c}{Non Participants} & \multicolumn{2}{c}{WSL Participants}  \\
% & Non\_Mean & non\_std & WSL\_mean & WSL\_std \\
 \hline
Indoor Area (m$^2$) & 185.5 & (81.3) & 199.9 & (85.0) \\
Lot Area (m$^2$)& 635.4 & (478.8) & 819.5 & (704.4) \\
Outdoor Area (m$^2$)& 462.1 & (435.4) & 617.11 & (658.5) \\
Rooms & 6.50 & (1.53) & 6.72 & (1.52) \\
Beds & 3.38 & (0.80) & 3.48 & (0.81) \\
Baths & 2.24 & (0.65) & 2.31 & (0.66) \\
Pool \% & 0.22 & (0.42) & 0.34 & (0.47) \\
Value (\$1,000)& 50.99 & (48.81) & 57.97 & (99.04) \\
Own \%* & 68.1 & (17.3) & 71.0 & (17.4) \\
\hline
N & \multicolumn{2}{c}{267,531}   & \multicolumn{2}{c}{24,206}   \\ \hline
\multicolumn{5}{l}{\footnotesize *ownership rates are inferred from block group level reporting}\\
\multicolumn{5}{l}{\footnotesize standard errors in parentheses}\\
\end{tabular}
\label{tab:descriptive stats}
\end{table}

\begin{figure}[htb]
   \centering
   \includegraphics[width=4in]{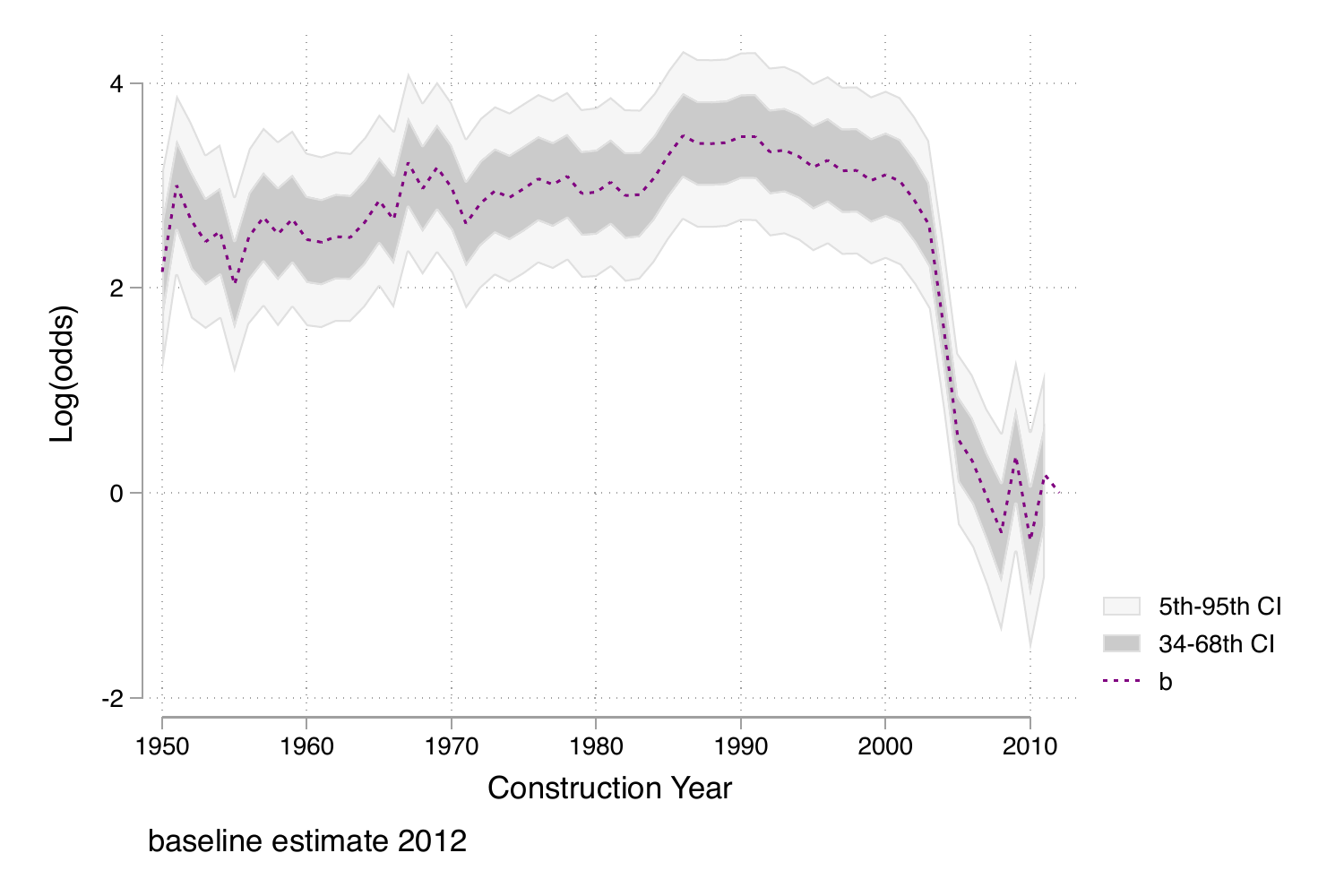}
   \caption{Estimated coefficients from a logistic regression with a dummy variable for each construction year, demonstrating that homes constructed after 2003 are much less likely to participate in the WSL program.}
   \label{fig:WSL logodds}
\end{figure}

Homes that were built after the 2003 turf restrictions had been enacted are much less likely to participate in the WSL program. Fig \ref{fig:WSL logodds} shows the results of a simple logistic regression exploring the conditional probability of WSL participation dependent on only a dummy variable for construction year, with 2012 as the baseline year. It is clear from this figure that the odds of WSL participation fall dramatically after about 2003, with no rebound yet apparent. A home has a 12\% probability of participating in WSL if it was built in 2003 or sooner, and only a 1\% chance of participating if it was built in 2004 or later.  Additionally, if a home constructed after 2003 did participate, they converted smaller areas, even though on average they had larger lot sizes.   This suggests that there may be a substantial, and growing, share of the Las Vegas housing stock which is ineligible for participation in the WSL program because there was no turf initially installed.  
 
%\begin{table}[htb]
%   \centering
%   \topcaption{Relationship between Lot Size and WSL conversion area}
%\begin{tabular}{l*{2}{r}}
%\hline\hline
%	&	Built in 2003 or Sooner	&	Built in 2004 or Later	\\
%   \hline
%Lot Area (m$^2$)	&	697.4	&	504.2	\\
%Lot Area if WSL = 1	&	818.1	&	860.8	\\
%WSL area if WSL = 1	&	121.2	&	61.1	\\
%\hline
%N	&	221,184	&	70,553	\\
%N if WSL = 1 & 23,415	&	791	\\
%\end{tabular}
%\label{tab:policy area}
%\end{table}

\section{Methods}

In this section, we explore the conditional probability of WSL adoption, describe the sampling method used to select focal neighborhoods and describe how ‘peers’ or neighbors are defined for each household within each neighborhood.  We next describe the main epidemic model, show how this model can be mapped into a hazard model, and finally discuss how we measure the epidemic effect. 

\subsection{Logistic Regression}

In order to describe the conditional probabilities of participation for homes with different physical characteristics, we use a simple logistic regression where the dependent variable is the log of the odds of WSL participation, and the independent variables are drawn from construction year, value, the presence of a swimming pool, lot size, outdoor area, and the size of the home. It is also important to point out that participating in the WSL program is a finite choice.  The terms of the WSL program prohibit homeowners from replacing rebated xeric landscaping with turf, a prohibition that is maintained through a deed restriction even after the home is sold.  There is a limited amount of turf on any property and once all turf has been removed, the home cannot participate further in the WSL program.\footnote{Roughly 10\% of homes in the WSL population had multiple approved conversions, removing some and then eventually all of the turf on the property.
In this study, homes with multiple conversions are excluded.}
As a result, homes that once chose to participate cannot do so after all the turf on the property has been removed, no matter how large other influential factors grow.
The directionality of WSL participation means that a hazard model rather than a logistic model is more appropriate for formally estimating the influence of various factors on WSL participation, and especially the existence of any potential peer effect.  

\subsection{Neighborhood Selection}\label{sec:neighselection}
Running an epidemic transmission model on the entire city of Las Vegas would be prohibitively expensive from a computational perspective, and so we randomly sample neighborhoods. 

Neighborhoods are selected by identifying sixteen randomly selected single family residential parcels (defined as seed parcels) within the complete dataset. A 1.5 km buffer is then drawn around that seed parcel, and all single family homes within this buffer zone are included in the core neighborhood for each seed parcel.  An additional 0.5 km buffer is drawn around each core neighborhood; homes in this buffer zone are included in the assessment of the characteristics of each homes' neighbors to avoid boundary problems in the identification of peers, but homes in the buffer zone are not analyzed as part of the neighborhood. Because the seed parcels are randomly selected, some neighborhoods overlap.  Homes that are contained in (for example) both seed neighborhood 1 and 3 are included in the analysis for both neighborhoods.  

\begin{table}
\begin{center}
   \topcaption{Summary statistics for all sample neighborhoods.}
\begin{tabular}{ c | r r l r l r l c c r } 
 \toprule
Seed & Homes & \multicolumn{2}{c}{Year Built}  & \multicolumn{2}{c}{Outdoor Area}& \multicolumn{2}{c}{Value (\$)}& Pool \% & Own \% & WSL	\\
\midrule
1 &3,344  & 1974.8 & (10.03) & 516.9 & (288.2) & 20,511 & (10,698)  & 0.18 & 0.55 &  163 	\\
2 &4,943  & 1991.3 & (2.50) & 457.4 & (187.6) & 66,196 & (27,539)  & 0.32 & 0.75 &  694 	\\
3 &5,353  & 1993.2 & (4.98) & 414.3 & (267.2) & 71,450 & (74,555)  & 0.33 & 0.64 &  708 	\\
4 &  6,034  & 1992.0 & (4.80) & 400.2 & (186.0) & 62,663 & (29,411)  & 0.34 & 0.68 &  798 	\\
5 &  5,011  & 1998.3 & (4.50) & 412.7 & (204.5) & 76,100 & (38,644)  & 0.30 & 0.66 &  521 	\\
6 &  4,070  & 1957.0 & (9.39) & 528.1 & (140.3) & 21,970 & (18,280)  & 0.23 & 0.59 &  300 	\\
7 &  3,607  & 2004.4 & (1.74) & 269.1 & (145.8) & 53,882 & (18,608)  & 0.09 & 0.52 &  126 	\\
8 &  699  & 2004.3 & (2.45) & 970.5 & (757.7) & 282,395 & (254,273)  & 0.60 & 0.89 &  49 	\\
9 &  5,984  & 2004.8 & (1.95) & 251.4 & (180.1) & 55,547 & (28,333)  & 0.12 & 0.60 &  78 	\\
10 &  4,882  & 1997.0 & (3.67) & 416.0 & (324.6) & 50,768 & (27,554)  & 0.28 & 0.79 &  466 	\\
11 &  1,950  & 1989.0 & (7.34) & 608.2 & (735.1) & 46,788 & (26,663)  & 0.26 & 0.66 &  266 	\\
12 &  2,645  & 1967.0 & (10.16) & 727.8 & (613.0) & 41,308 & (37,481)  & 0.43 & 0.64 &  288 	\\
13 &  3,433  & 1997.4 & (6.40) & 552.2 & (583.8) & 52,410 & (20,682)  & 0.28 & 0.86 &  509 	\\
14 &  1,839  & 2008.1 & (1.84) & 254.9 & (99.9) & 50,241 & (21,167)  & 0.05 & 0.86 &  9 	\\
15 &  6,316  & 1997.1 & (4.77) & 396.5 & (232.7) & 52,411 & (19,209)  & 0.22 & 0.71 &  843 	\\
16 &  7,033  & 2001.9 & (2.98) & 264.8 & (144.2) & 45,457 & (15,391)  & 0.15 & 0.62 &  360 	\\
\midrule
All  &  61,385  & 1993.0 & (14.42) & 415.8 & (342.2) &  54,827 & (51,120)  & 0.23 & 0.67 &  5,431\\
City & 291,737  & 1992.1 & (14.68) & 475.0 & (460.1) &  51,566 & (54,789)  & 0.23 & 0.68 &24,206\\
 \bottomrule
\multicolumn{5}{l}{\footnotesize Standard errors in parentheses}\\
\end{tabular}
\end{center}
\end{table}

\subsection{Network and Neighborhood Definition}\label{sec:network}
\begin{figure}
  \centering
      \includegraphics[width=0.5\textwidth]{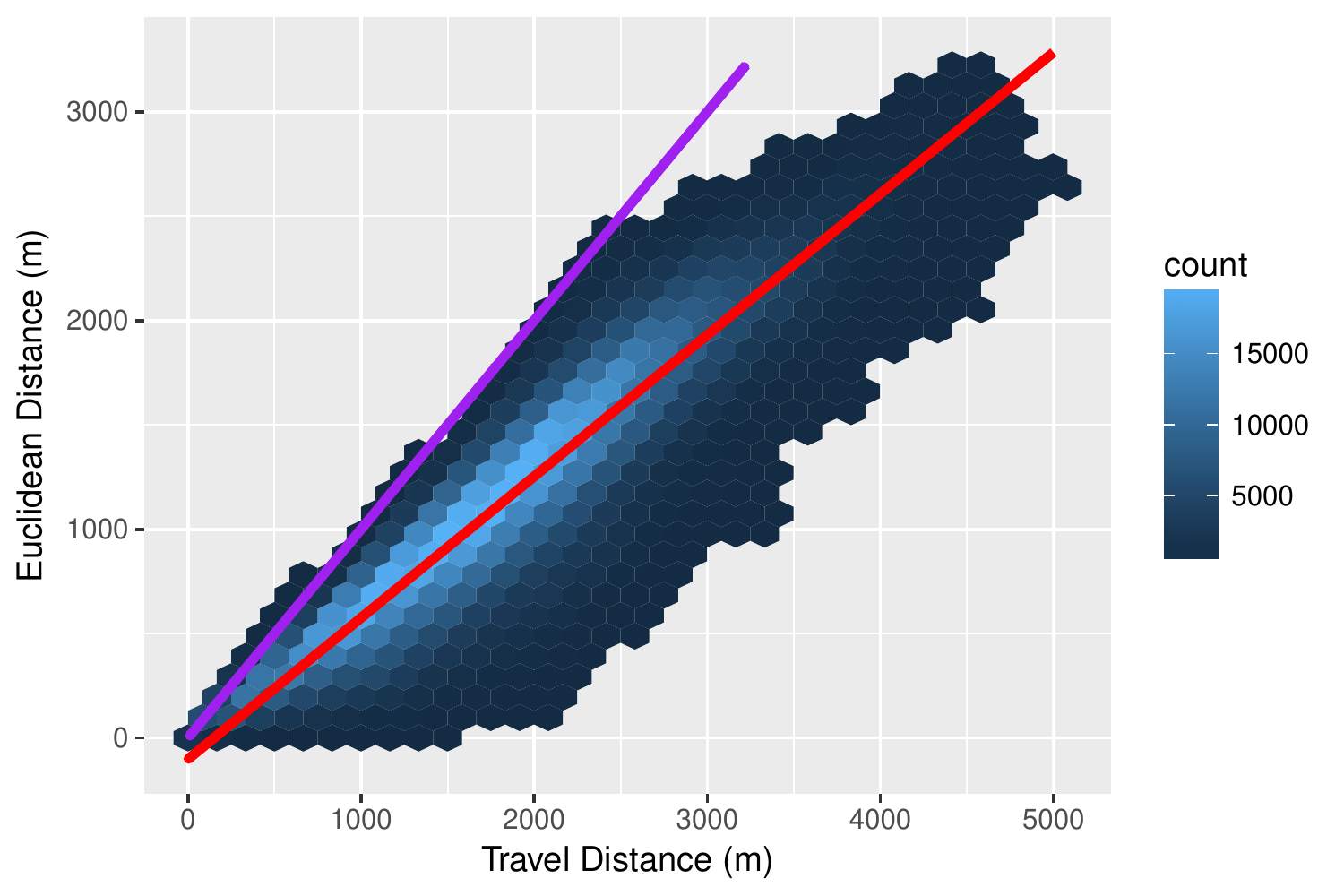}
  \caption{The purple line shows the line with slope 1. Red shows the best fit line, slope 0.7.  Note that there is substantial variability in the mapping from parcel to parcel euclidean distance and on-road travel distance. }
  \label{fig:network distance}
\end{figure}

While some models used to study peer effects consider \textit{cumulative} count data at ZIP-code and street level \cite{bollinger_peer_2012,richter2013social} and within a certain euclidean distance \cite{graziano_spatial_2015}, here we fully exploit the rich spatial detail in our dataset and consider a more refined network structure which allows to model the probability of program participation at a an \textit{individual household} level. A similar approach has been considered in \cite{towe_contagion_2013}, where they build a network of neighbors interactions based on the first $k$ nearest neighbors within a radius of euclidean distance.  
There are several methods available for defining exactly which nearby pairs of houses qualify as \textit{neighbors}.  
Here we define the distance between a pair of homes in two different ways:  the euclidean distance $d_e$ between the two homes, and the on-road travel distance $d_n$ between the same two homes. 
On-road travel distance is calculated by creating a spatially embedded network of all parcels and roads in a given neighborhood. 
Fig \ref{fig:network distance} compares the on-road travel distance to Euclidean distance for all pairs of homes in one Las Vegas neighborhood. 
We then select a threshold $\tau_d$, and two homes in a given neighborhood are defined \textit{neighbors} if the distance between them is less than the selected threshold $\tau_d$. Formally, this is represented by creating an unweighted edge between the two houses. 
We apply three thresholds: 0.1, 0.2 and 0.3 km to the two different distance measures to create six different networks on which to test the existence of a peer effect. 

We use both euclidean distance and on-network travel distance in these models because euclidean distance is the standard approach and computationally simpler, but on road travel distance is a more relevant measure of likely peers in Las Vegas’ car centered urban layout.  
Homes that are physically nearby but on different streets, for example `back yard neighbors’, may have very little potential for interaction, and thus little potential for influence. 

\subsection{Epidemic Model} \label{sec:epidemicmodel}
We mathematically model the dynamics of the participation in the WSL program as a discrete-time epidemic spreading model (SEIR) with \textit{autoinfection}. In contrast to standard epidemic models, we allow a house to activate because of a combination of \textit{endemic} causes in addition to the standard \textit{epidemic} transmission. 

In contrast to standard economic survival models, this mechanistic network approach allows for a straightforward representation of peer effects by accounting for an explicit and direct interaction between neighboring houses by means of an epidemic transmission parameter and a well defined network structure. Nodes in the network are houses and the set of all nodes is denoted by $V$. Edges are created between neighboring houses when the distance between the two homes is less than the distance threshold selected as described in section \ref{sec:network}.

Finally, in contrast to the methods studied in  \cite{young_innovation_2009} in the context of innovation spreading, where they assume that members of a group interact more or less at random, in our case information flows between individuals in a fixed network and the dynamics are explicitly influenced by the specific network topology considered.

We assume that a house $i$ can be in any of four states $x_{i}^{t}\in \{S, E, I, R\}$ representing susceptible, exposed, infected and recovered respectively. The allowed transitions in state are $S\rightarrow E$, $E\rightarrow I$ and $I\rightarrow R$ and once a node is recovered it will remain in that state forever (irreversible dynamics). 
A susceptible house $i$ can become exposed with a probability $\alpha$ whenever one of its neighbors $j$ is infected. This captures the peer effect that results from a WSL participating household influencing a nearby non-participating household. Differently from a standard epidemic model  \cite{kermack1932contributions,anderson1992infectious}, we allow households to participate in the program \textit{independently} from their neighbors with a probability $\mu_i^t$ at each time step $t$, and we call this \textit{autoinfection probability}, which represents the endemic component. This assumption allows us to capture the possibility that a household chose to participate because of a global effect that affects all houses equally, because of a house-dependent local effect that affects each houses differently based on a set of covariates, or any other reason independent of neighborhood based peer effects.  Examples of the global effect are city-wide marketing campaigns, or peer effects through friends and colleagues who are not neighbors.  The house-dependent local effect includes covariates like house value, construction year, and outdoor area that are associated with the decision to participate.
%In our application, houses that do not participate in the WSL program are susceptible; houses that participate are considered exposed during the time window from when they file the application to when this is approved and the rebate is given to them by the water agency; from this latter date, houses are considered infectious, because their participation can be seen by their neighbors by direct observation of their yards. In addition, we assume that the influence that a participant can exercise on their neighbors eventually fades out, thus allowing for a recovery state after which past WSL participation no longer influences their neighbors' decisions about WSL participation. 

In addition to considering finite recovery periods, we also consider a recovery period of infinite length, which in practice means that a WSL participating homes' influence on their neighbors never fades, and is equivalent to the SEI model in epidemiology.

In summary: we assume that a house can chose to participate in WSL (become active) by either \textit{autoinfection} or contagion from its active neighbors. Calling $\mu$ the $N$-dimensional vector with $i$-th entry $\mu_{i}^t$, the probability the house $i$ autoinfects at time $t$, and $\alpha$ the \textit{transmission probability} from house $i$ to house $j$, we can write the probability of the state of a house at a given time step $x_{i}^{t}$. 
In principle the transmission parameter could depend both on time and on the combination $(i,j)$ of houses involved. However, this considerably increases the number of parameters and thus the risk of overfitting. Moreover, we do not have any prior knowledge about pairwise relationship between houses that would support the idea of treating them differently. Therefore in this work we assume that $\alpha$ is constant in time and does not depend on what pairs of houses we consider (whereas the autoinfection probability $\mu_i^t$ differs from house to house). 
The duration of the exposed period $E \rightarrow I$ is given as measured for each household.
For simplicity we also assume that an infected house switches to recovered after a fixed period of time $\tau_R$ from the time it first became infective $t_{iI}$, with the latter quantity depending on the house considered.
Notice that this implies that the only stochastic quantity is the activation time: the time of switching $S \rightarrow E$. 
We can write the probability of a house $i$ activating at time $t_{iE}$, given its initial state $x_i^0$ and the state of its neighbors in time as: 
\bea
P(t_{iE}=t_i>0| \alpha,\mu)&=& \mathbb{I}_{x_{i}^{0}=S} \, \rup{\prod_{t=1}^{t_{i}-1} P(x_{i}^{t}=S|\alpha,\mu) } \, P(S\rightarrow E, t_i|\alpha,\mu) \nonumber \\
&=&\mathbb\mathbb{I}_{x_{i}^{0}=S} \, \rup{ \prod_{t=1}^{t_{i}-1} (1-\mu_i^t) \,   \prod_{k \in \partial i | t_{kI}<t_{i}-1} (1-\alpha)^{t_{i}-\tau_{kiI}-1}} \nonumber \\
&&\times \rup{1-(1-\mu_i^{t_i})(1-\alpha)^{n_{i}}}  \label{poft}
\eea
for $t_{i} \in [0, T]$, where $n_{i}$ is the number of neighbors of $i$ that are infected at time $t_i$ and $\tau_{kiI}$ is the number of time steps that neighbor $k$ had to infect house $i$, i.e. $\tau_{kiI}= t_{kI}$ if $t_{kI}+\tau_R >t_i$, $\tau_{kiI}= \tau_R$ if $t_{kI}+\tau_R \leq t_i$; $\partial i$ denotes the set of neighbors of $i$. 

Unlike the parameter estimates from a hazard model (as in \citeauthor{towe_contagion_2013}), $\alpha$ is directly interpretable as the transmission probability: the probability that a non-participating home will be `infected' in a given month by one of their infectious neighbors, and is bounded between zero and one. Outside of corner cases, the probability that a non-participating home will be infected by this neighbor is then $(1-(1-\alpha)^{\tau_R})$, from the last component of (\ref{poft}).

For a house that never activates before time $T$, we define $t_{iE}=\infty$ and its trajectory in time has probability:
\bea 
P(t_{iE}=\infty | \alpha,\mu)&=&\mathbb{I}_{x_{i}^{0}=S}\,\prod_{t=1}^{T} P(x_{i}^{t}=S|\alpha,\mu) \nonumber \\
&=&\mathbb{I}_{x_{i}^{0}=S} \,\prod_{t=1}^{T} (1-\mu_i^t) \,   \prod_{k \in \partial i | t_{kI}<T} (1-\alpha)^{T-\tau_{kiI}} \label{nonactive}
\eea

The joint probability of all houses' trajectories is represented as $P(\bar{t}|\alpha,\mu)=P({t}_{1},\dots,{t}_{N}|\alpha,\mu)$ and is factorized on the nodes as:
\be \label{eq:totP}
 P(\bar{t}|\alpha,\mu)=\prod_{i \in V}P(t_{iE}|\alpha,\mu)
\ee
and the log likelihood can be derived by inserting (\ref{poft}) and (\ref{nonactive}) into (\ref{eq:totP}) and taking the logarithm (see appendix \ref{apx:logL}). 
Notice that other versions of epidemic models are easily obtained from this general formulation. For instance, an $SI$ model is obtained by setting $\tau_{iE}=t_{iI}-t_{iE}=0$ and $\tau_R=\infty$.

Notice that $t_{iE}$ are the observed data, while $\alpha,\mu$ are parameters that need to be estimated. Our goal here is to estimate the value of the transmission parameter $\alpha$ and see if it shows evidence of peer effects. We define evidence of a peer effect through two steps:  i) if we find that $\alpha=0$ according to our model, this is interpreted as no peer effect because the transmission probability is zero; ii) if we find that $\alpha>0$, then we compare this model with the model where $\alpha$ is fixed at zero. If the former is better (see Section \ref{sec:measuring} for the model selection criteria we use) then we interpret this result as evidence that the peer effect is present: the model that has a non-zero transmission probability better fits the observed data than a model where no peer effect is included.

\subsection{Mapping the Epidemic Model into a Hazard Model}\label{sec:mapping}
A common model used in economic, marketing or sociology literature to describe contagion dynamics is the hazard model, as used, for instance, to study similar problems in \cite{bollinger_peer_2012,iyengar2011opinion,aral2012identifying,towe_contagion_2013}. An additive-multiplicative regression model has also been introduced in epidemiology \cite{hohle2005inference,hohle2008spatio,hohle2009additive} to study the spread of diseases along with an R package \cite{JSSv077i11}. However a clear mapping between the mechanistic epidemic modeling described in Section \ref{sec:epidemicmodel} and hazard models as defined in the context of survival analysis is still lacking. Here we propose such mapping showing (see appendix \ref{apx:epidemichazard}) that the discrete-time versions of the additive-multiplicative hazard model presented in \cite{hohle2008spatio,hohle2009additive} using the formalism of counting processes is equivalent to the mechanistic discrete-time SEIR model with autoinfection described above when:
\be \label{eq:epihaz}
\log \, \rup{- \log \bup{1-\mu_{i}^t} }=\bar{x}_{i} \cdot \bar{\beta} +\log \, \lambda_{0}^t  
\ee
where $\lambda_0^t$ is the baseline hazard at time $t$ which represents a global contribution to the probability of getting infected which is the same for all houses; $\bar{x}_i$ is a vector of covariates and $\bar{\beta}$ is a vector of parameters coupling the covariates, in a similar way as in regression analysis. The model in \cite{hohle2008spatio,hohle2009additive} does not consider event ties, i.e. activation events happening at the same time. This is due to the formalism of counting processes that does not account for them, although in principle one can avoid them by adding a random and small perturbation to the events' times so to resolve ties. In our problem we have access to the exact program's application date (day, month and year). Although this does not completely rule out ties, two houses could potentially file the same day, the precise day of filing might not be as precisely determined than the month of application. In fact the application date can be influenced by delays in delivering the documents or in the administrative time to process it. In other words, the exact day  might be affected by an error of several days. To account for this we use as time step in describing the dynamical process a unit of one month (delays of months in the previous two factors are much less likely) so that we count as exposure time $t_{iE}$ the month when the application was filed. With this time unit we are subject to event ties, but our discrete time mechanistic epidemic model is not affected by it because we consider parallel updates (as opposed to sequential): all the network's nodes update at the same time, given the network state at the previous time step.

From (\ref{eq:epihaz}) we can see that the autoinfection probability $\mu_i^t$ of the epidemic model is mapped into a combination of a (time-dependent) global contribution represented by the baseline hazard and a node-dependent contribution due to the node covariates $\bar{x}_i$ as in a regression framework.
This mapping is allowed by the additive way through which the epidemic and endemic terms are coupled together. A similar straightforward mapping between epidemic and hazard model cannot be obtained in the case of a multiplicative hazard model (as used in \cite{towe_contagion_2013}), because in that case these two terms cannot be trivially decoupled (see appendix \ref{apx:epidemichazard}).
In fact a purely multiplicative model has the property that if either one of the epidemic or endemic effect is zero, than the overall hazard rate is zero, regardless the other effect. This is not the case in models where the hazard rate is instead a superposition of epidemic and endemic effects, as in our case or for additive-multiplicative survival models.

In this work we consider as house covariates: the build year, outdoor area, value, percentage of house ownership in the block, presence of a pool and a dummy variable indicating whether or not the house was built before 2003: These are in fact the main covariates influencing the probability of participation resulting from logistic regression (see Section \ref{sec:lrResulrs}).  
\subsection{Inference}
Now we turn our attention to estimating the value of $\alpha$ which we will use to assess the presence of peer effects, as explained above. We use maximum likelihood inference on the epidemic model's likelihood integrated with house level covariates. This can be obtained by combining (\ref{eq:totP}) with the mapping $(\ref{eq:epihaz})$. This allows us to capture household level variations in autoinfection probability and thus control for heterogeneity among individual home characteristics.
%\CBcom{in addition to exploring the value of $\alpha$ as described above, we observe the temporal dynamics of WSL participation rates in each neighborhood.  Homophliy or CO driven influences in WSL participation could lead to the type of spatial clustering that we observe, but the temporal patterns should be distinguishable. in this section, we explore the temporal patterns. Basically, this kind of inferrence analysis is useful for.... because .... }

By doing this, the log likelihood of the system can be written as a function of the covariates as:

\bea \label{eq:logLHM}
\mathcal{L}(\bar{t}|\lambda_0^t,\bar{\beta},\alpha)&=&\sum_{i \in V | \bar{x}_i^T=S} \rup{ -\sum_{t=1}^{T} \lambda_{0}^t \,e^{\bar{x}_{i}\cdot \bar{\beta}} + \sum_{k \in \partial i | t_{kI}<T} ({T-t_{kiI}}) \log (1-\alpha)} +\\
&&+\sum_{i \in V | \bar{x}_i^T\neq S} \left[  \sum_{k \in \partial i | t_{kI}<t_{iE}-1} ({t_{iE}-\tau_{kiI}-1}) \log (1-\alpha) +\log \rup{1-\bup{e^{-\lambda_{0}^t}}^{e^{\bar{x}_{i} \cdot \bar{\beta}}}(1-\alpha)^{n_{i}}} \right. \nonumber \\
&& \left.-\sum_{t=1}^{t_{iE}-1} \lambda_{0}^t \,e^{\bar{x}_{i}\cdot \bar{\beta}}\right]  \nonumber
\eea

The maximum likelihood estimator $\hat{\Theta}=(\lambda_0^t,\bar{\beta},\alpha) = arg \sup_\Theta \mathcal{L}(\bar{t}|\lambda_0^t,\bar{\beta},\alpha) $ is a complicated polynomial, and it is difficult to analytically derive the derivatives with respect to the parameters, so we use standard bounded numerical optimization methods as L-BFGS-B \cite{byrd1995limited} with constraints on the parameters such that $0 \leq \alpha \leq 1$ and $0 \leq \bup{e^{-\lambda_{0}^t}}^{e^{\bar{x}_{i} \cdot \bar{\beta}}}\leq 1$.

%With this mapping we can then estimate the parameters $(\beta,\lambda_0^t)$ through maximum likelihood estimation by inserting the equation $\log \bup{1-\mu_{i}^t} =-\lambda_{0}^t \,e^{x_{i} \cdot \beta }$ into the log likelihood expression (\ref{eq:logL}).
Here we consider a piece-wise constant baseline hazard $\lambda_0^t$ which is constant within three intervals defined by the rebate price windows in months: [Jan 2004, Dec 2006], [Jan 2007, Dec 2007] and [Jan 2008, Dec 2015] as shown in Fig. \ref{fig:WSL_summary}. These price changes were also temporally consistent with other major environmental changes in the Las Vegas metro area.  The only sustained marketing campaign for WSL participation that SNWA staff have reported to us occurred in spring and fall of 2007, roughly coincident with the highest rebate price.  Similarly, the price change in Jan 2008 occurred at the same time as the most dramatic effects of the 2008 recession were becoming apparent. Thus, we believe these breakpoints are good candidates for capturing global effects.
The estimated standard errors of the parameters are obtained from the relation with the Fisher information matrix \cite{pawitan2001all}, which is derived by numerically calculating the Hessian of the log likelihood function (\ref{eq:logLHM}).

\subsection{Measuring the Epidemic Effect}\label{sec:measuring}
To test whether an epidemic contribution in the dynamics of adoption exists we adopt two different strategies. First, from an \textit{inference} perspective, the model with lower Akaike Information Criterion (AIC) \cite{akaike} between the case where there is no peer effect, $\alpha=0$, and the case where the peer effect is present, $\alpha>0$ is more likely to be correct.
Secondly, from a \textit{prediction} perspective we select the model the has higher predictive performance. Notice that, arguably counter intuitively, these two strategies may give different results as shown in different contexts in \cite{shao1993linear,valles2017consistency}. 
In both the inference and prediction models, we use a spatially accurate network structure, that does not induce any boundary conditions.  
Our break between the exposed and infectious stages of the SEIR model handles challenges associated with reflection.  
The covariates regarding home characteristics, which are wrapped into the estimate of $\mu_i$, manage some homophily and correlated unobservables problems. 
Additionally, in the prediction analysis, we capitalize on the fact that correlated unobservable characteristics about physical neighborhood characteristics and the neighborhoods residents are likely to change slowly relative to the WSL adoption rate, and so the temporal dynamics of adoption can be a strong signal for the existence of peer effects.  When both prediction and inference methods suggest the existence of peer effects, we believe that this is a strong and conservative signal that the effect exists in a meaningful way.

\section{Results}
  
\subsection{Conditional Probabilities of Participation}\label{sec:lrResulrs} 

Using a very simple logistic regression to explore the characteristics that are associated with eventual WSL participation, we find that across the city as a whole, whether the home was built before or after 2003 is the single best available determinant of WSL participation. 
The presence of a pool is the next most significant factor associated with WSL participation.  
Larger in lot areas and indoor areas are associated with a small increased likelihood of WSL participation, while (after controlling for size) increased value is associated with a decreased likelihood of participation. Tab. \ref{tab:logistic} shows results.  
We find that for the typical house without a pool, the probability of participating in the WSL program is 0.07.  
The probability of a house participating in the WSL program if they do have a pool is 0.12.  
The probability of a house participating in the WSL program if it was built before 2003 and has a pool is 0.14, but if it was built after 2003 and doesn't have a pool, it's only 0.01. 
Adding various physical descriptors of the home and property (Lot Area, Value, and Indoor Area) only has a small influence on the conditional probability of these two major factors: for the typical pre-2003 house with a pool, decreasing the value by \$10,000, combined with increasing the lot area and indoor area by 10 m$^2$ each only changes the conditional probability from 0.119 to 0.122, from column (3). 
While statistically significant, these differences are not important in practice.  In column (4) of Tab. \ref{tab:logistic}, we demonstrate that separating homes into those constructed before and after the 2003 turf restrictions occurred is more consistent with the structure of the data than using dummy variables by individual decades of construction. 
Fig \ref{fig:WSL logodds} shows the coefficients and standard errors for this regression if a dummy variable is included for each possible construction year. 
This further demonstrates the empirical support for the 2003 split in the effect of home construction year on WSL participation. 
Column (5) shows the effect of including spatial fixed effects at the census tract level, which again somewhat decreases the marginal effect of pools and the effect of Post-2003 construction.  
These results motivates the choice of covariates used in fitting the epidemic model of Section \ref{sec:mapping}.

\begin{table}[htb]
   \centering
   \topcaption{Logistic regressions exploring conditional probability of participation in WSL during the study period. Lot Area, Value, Year Built, and Indoor Area are normalized to 0, so that the constant represents probability of participation of the `typical', pre-2003 house, which works out to roughly a 7\% probability.}
\def\sym#1{\ifmmode^{#1}\else\(^{#1}\)\fi}
\begin{tabular}{l*{5}{c}}
\toprule
&\multicolumn{1}{c}{(1)}&\multicolumn{1}{c}{(2)}&\multicolumn{1}{c}{(3)}&\multicolumn{1}{c}{(4)}&\multicolumn{1}{c}{(5)}\\
                    &\multicolumn{1}{c}{Pool}&\multicolumn{1}{c}{+Turf Policy}&\multicolumn{1}{c}{+Home}&\multicolumn{1}{c}{+Build Year}&\multicolumn{1}{c}{+Tracts}\\
\midrule
WSL                 &                     &                     &                     &                     &                     \\
Has Pool            &       0.599\sym{***}&       0.422\sym{***}&       0.225\sym{***}&       0.305\sym{***}&       0.139\sym{***}\\
                    &    (0.0143)         &    (0.0145)         &    (0.0160)         &    (0.0162)         &    (0.0165)         \\
[1em]
Built after 2003    &                     &      -2.278\sym{***}&      -2.294\sym{***}&                     &      -1.987\sym{***}\\
                    &                     &    (0.0365)         &    (0.0371)         &                     &    (0.0452)         \\
[1em]
Value (\$10,000)    &                     &                     &    -0.00985\sym{***}&     -0.0179\sym{***}&     -0.0102\sym{***}\\
                    &                     &                     &   (0.00211)         &   (0.00163)         &   (0.00169)         \\
[1em]
Lot Area (10m$^2$)  &                     &                     &     0.00271\sym{***}&     0.00347\sym{***}&     0.00221\sym{***}\\
                    &                     &                     &  (0.000129)         &  (0.000135)         &  (0.000157)         \\
[1em]
Indoor Area (10m$^2$)&                     &                     &      0.0182\sym{***}&      0.0166\sym{***}&     0.00980\sym{***}\\
                    &                     &                     &   (0.00137)         &   (0.00121)         &   (0.00130)         \\
[1em]
Construction Decade FE & No & No & No & Yes & No \\
Tract level FE & No & No & No & No & Yes \\
[1em]
Constant            &      -2.569\sym{***}&      -2.261\sym{***}&      -2.227\sym{***}&      -5.504\sym{***}&      -2.263\sym{***}\\
                    &   (0.00821)         &   (0.00846)         &   (0.00864)         &     (0.243)         &     (0.227)         \\
\midrule
Observations        &      291737         &      291737         &      290188         &      290188         &      289950         \\
Pseudo \(R^{2}\)    &       0.010         &       0.057         &       0.063         &       0.045         &       0.093         \\

\multicolumn{6}{l}{\footnotesize Standard errors in parentheses}\\
\multicolumn{6}{l}{\footnotesize \sym{*} \(p<0.05\), \sym{**} \(p<0.01\), \sym{***} \(p<0.001\)}\\
\end{tabular}
\label{tab:logistic}
\end{table}

\subsection{Inference Performance}

In Tab.~\ref{table:inference} we show the inference results for three models of each neighborhood, one for each of the different methods for selecting potential peers: i) using euclidean travel distance between pairs of homes, ii) using on-road travel distance between homes, and iii) the purely endemic model with no peers at all.  Two further parameters need to be selected: recovery time and distance threshold ($\tau_R,\tau_d$). 

In order to parsimoniously present results, we select $\tau_R =12$ (in months) as the preferred $\tau_R$ value. The results presented in Tab.~\ref{table:inference} reflect this value with additional values of $\tau_R$ presented in appendix Tab. \ref{table:inference_recovery}.
This recovery period was selected after testing values $\tau_R \in \{ 1,4,6,12, \infty\}$, and finding that in general, $\tau_R =12$ has the lowest AIC values among the models we tested.  
We find that the results for $\tau_R \in \{ 4,6\}$ are broadly consistent with the results found when $\tau_R = 12$.  
They have a statistically significant $\alpha>0$ and a lower AIC than the model in which peer effects are excluded ($\alpha=0$).  For $\tau_R \in \{ 1,\infty\}$ we find that $\alpha$ is more often indistinguishable from zero and AIC values are higher, demonstrating that models with a measurable peer effect fit the data better, and the temporal duration of that peer effect is about a year.\footnote{Note that we are selecting preferred models on the basis of a better model fit, \textit{not} a positive $\alpha$ value, but the model parameters that lead to a positive $\alpha$ value also fit the data better than models where the selected parameters result in $\alpha = 0$.}
One possible explanation for this is that the cases $\tau_R \in \{ 1,\infty\}$ are two extremes: either the recovery is too fast and thus it does not allow for enough time for the contagion to happen or it never ends, so that neighbors who had activated years ago exercise the same peer effect as recently activated neighbors. 
The latter case represents the standard SEI model, where an active individual is always infectious. 
Our finding that an SEIR model with $\tau_R \in \{ 4,6,12\}$ is preferred to an SEI model (at least in terms of AIC) suggests that contagion effect fades out with time. \cite{graziano_spatial_2015} obtained similar results on the duration of influence when analyzing contagion effects in solar panel installations.  

There is less consistency across neighborhoods in model response to $\tau_d$.  We test values of $\tau_d\in \{0.1,0.2,0.3 \}$ (in km), and for each individual neighborhood, select the $\tau_d$ value with the lowest AIC value. In general, the choice of $\tau_d\in \{0.1,0.2,0.3 \}$ does not have a large effect on $\alpha$ results or model fit. 
It makes sense to allow the distance threshold for allowing peer effects to vary by neighborhood, as there is significant variance in home density across our sample neighborhoods. 
We might expect neighborhoods with small lot sizes to have a shorter distance threshold for peer effects than neighborhoods with large lot sizes because so many more homes will be included in that range. Exhaustive combinations of models with different ($\tau_R,\tau_d$) parameter choices are available from the authors upon request.  Other constant values or functional forms could have been considered for both the recovery time and distance thresholds, but the values we have chosen are reasonable, consistent with the literature, and finding the best values of these two parameters is beyond the scope of this paper. 

\begin{figure}[htb]
   \centering
   \includegraphics[width=6in,trim={0.5cm 1.2cm 0.5cm 0.5cm}, clip]{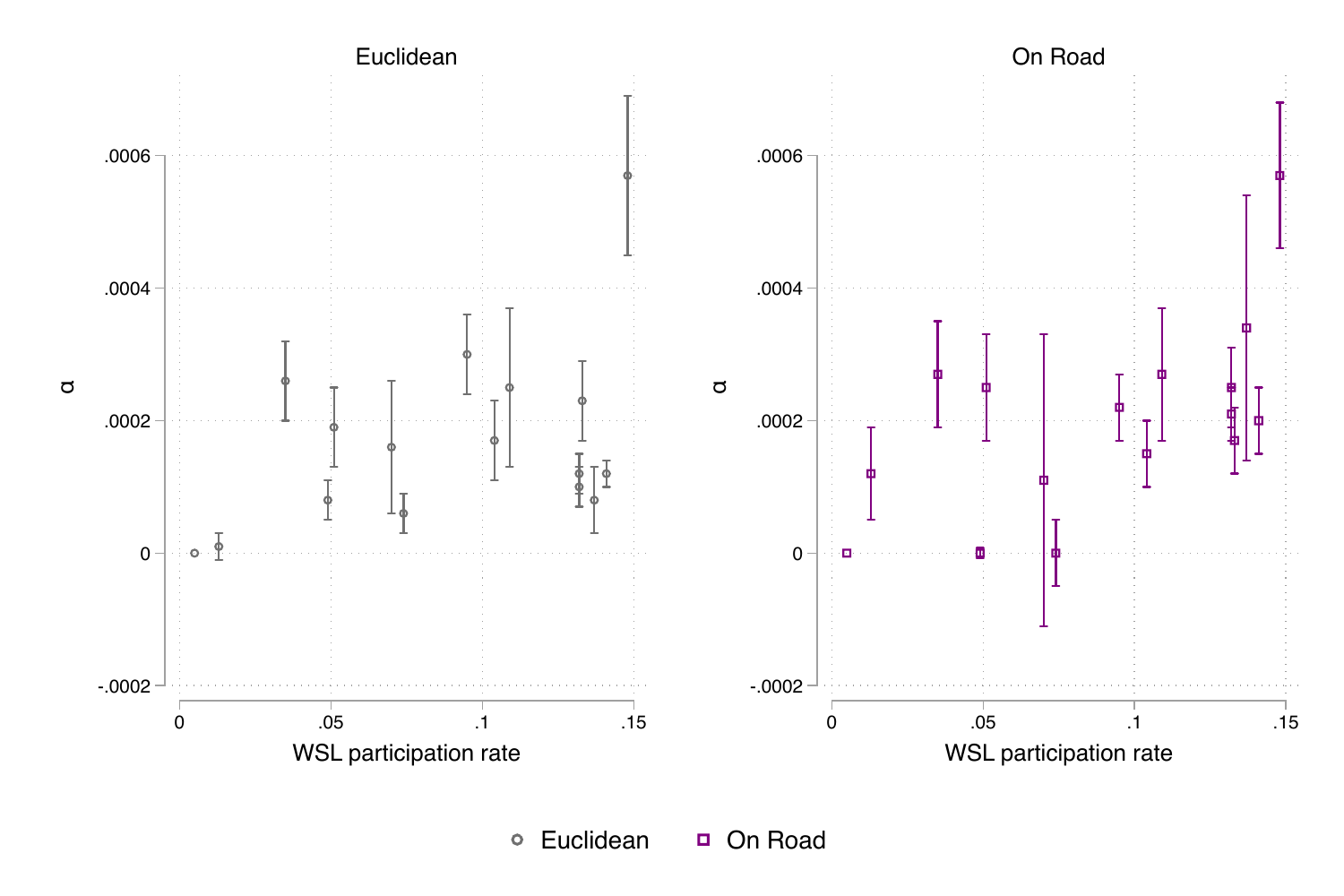}
   \caption{Estimated $\alpha$ values for Euclidean and On Road models presented in Tab. \ref{table:inference}, compared to neighborhood level WSL participation rates. Standard errors for neighborhood 14 ($\alpha = 0$ with only 9 WSL participating homes) are not displayed, but would extend well beyond the range of the y-axis shown.}
   \label{fig:MLE results}
\end{figure}

Fig. \ref{fig:MLE results} summarizes Tab. \ref{table:inference}, comparing the estimated $\alpha$ values to the WSL participation rate in each neighborhood.  From this figure, it is clear that the majority of the neighborhoods show a statistically significant $\alpha >0$, and that as WSL participation rates rise within each neighborhood, so too does the magnitude of the peer effect, for both the Euclidean and On Road distance measures. Error bars show the standard error. 

Neighborhoods $n\in\{2,3,4,5,7,10,13,15,16 \}$ show statistically significant and positive $\alpha$ estimates at the 1\% level for both the euclidean and on-network distance thresholds. Neighborhoods $n\in\{11,12\} $ show positive $\alpha$ estimates that are only weakly significant (at the 10\% level or greater for both distance measures).  
Finally, neighborhoods $n\in\{1,6,8,9,14\} $ either don't have an epidemic effect or it is not significant.  
Neighborhoods 8, 9 and 14 have the lowest counts of WSL participating homes, at 49, 78 and 9 homes respectively, and also have a median construction year after 2004, meaning that the majority of homes were constructed with significant turf restrictions in place, a policy that may effectively put a home in the recovered status from its construction onwards.  These neighborhoods simply may not have a large enough population of WSL participants for a peer effect to matter.
The other two neighborhoods with no measurable epidemic effect, $n\in\{1,6\}$ show the lowest average assessed home value (around \$21,000, relative to a citywide mean of \$51,000) which is about half the average value of the next lowest valued neighborhood in our sample.  
Because many neighborhoods in Las Vegas show convincing evidence of a peer effect, its absence in some places should not be taken as evidence that peer influence has no effect on WSL participation rates.  
 This method has controlled for reflection challenges through use of an SEIR model, and addressed some challenges with homophily and correlated unobservables through careful selection of control variables and use of fixed effects. 
These results provide strong suggestive evidence of peer effects in WSL participation at the neighborhood level.

\begin{table}[!htbp]\label{table:inference}
\def\sym#1{\ifmmode^{#1}\else\(^{#1}\)\fi}
\centering
\caption{Maximum likelihood inference results. The recovery time is $\tau_R=12$ months; $\alpha$ is the epidemic transmission parameter and $STE_{\alpha}$ its standard error; $\tau_d$ is the distance threshold used to build the network;  WSL rate is the eventual WSL participation rate. Within each neighborhood's results, models are ordered by AIC, a measure of model fit.}
%\begin{tabular}{lllllllrll}
\begin{tabular}{rlrlrrrr}
\toprule
 Neighborhood & Distance Metric &      AIC &     $\alpha$ &  $STE_{\alpha}$ &  $\tau_d$ &  WSL rate& No. Homes \\
\midrule
1	&	   euclidean 	&	2855	&	    0.00008\sym{*} 	&	0.00003	&	0.3	&	0.049	&	3344	\\
	&	no peer	&	2865	&	             	&	    	&	       	&		&		\\
	&	  on-road 	&	2867	&	0	&	0.000008	&	0.1	&		&		\\
\addlinespace[0.5ex]
2	&	   euclidean 	&	10786	&	  0.00012\sym{***} 	&	0.00002	&	0.3	&	0.141	&	4931	\\
	&	  on-road 	&	10813	&	   0.00020\sym{***} 	&	0.00005	&	0.3	&		&		\\
	&	no peer	&	10835	&	             	&	      	&	     	&		&		\\
\addlinespace[0.5ex]
3	&	  on-road 	&	10967	&	  0.00021\sym{***} 	&	0.00004	&	0.3	&	0.132	&	5340	\\
	&	   euclidean 	&	10977	&	  0.00012\sym{***} 	&	0.00003	&	0.2	&		&		\\
	&	no peer	&	11040	&	             	&	      	&	     	&		&		\\
\addlinespace[0.5ex]
4	&	  on-road 	&	12412	&	  0.00025\sym{***} 	&	0.00006	&	0.2	&	0.132	&	6027	\\
	&	   euclidean 	&	12415	&	  0.00010\sym{***} 	&	0.00003	&	0.2	&		&		\\
	&	no peer	&	12471	&	             	&	     	&		&		&		\\
\addlinespace[0.5ex]
5	&	   euclidean 	&	8402	&	   0.00017\sym{**} 	&	0.00006	&	0.1	&	0.104	&	5003	\\
	&	  on-road 	&	8415	&	   0.00015\sym{**} 	&	0.00005	&	0.3	&		&		\\
	&	no peer	&	8426	&	             	&	    	&		&		&		\\
\addlinespace[0.5ex]
6	&	   euclidean 	&	5087	&	    0.00006\sym{*} 	&	0.00003	&	0.3	&	0.074	&	4069	\\
	&	no peer	&	5091	&	            	&	  	&		&		&		\\
	&	  on-road 	&	5093	&	0	&	0.00005	&	0.3	&		&		\\
\addlinespace[0.5ex]
7	&	  on-road 	&	2211	&	  0.00027\sym{***} 	&	0.00008	&	0.3	&	0.035	&	3603	\\
	&	   euclidean 	&	2221	&	  0.00026\sym{***} 	&	0.00006	&	0.2	&		&		\\
	&	no peer	&	2328	&	            	&	      	&		&		&		\\
\addlinespace[0.5ex]
8	&	   euclidean 	&	835	&	0.00016	&	0.0001	&	0.3	&	0.07	&	696	\\
	&	  on-road 	&	835	&	0.00011	&	0.00022	&	0.2	&		&		\\
	&	no peer	&	836	&	            	&	     	&		&		&		\\
\addlinespace[0.5ex]
9	&	  on-road 	&	1463	&	    0.00012\sym{\circ} 	&	0.00007	&	0.3	&	0.013	&	5980	\\
	&	   euclidean 	&	1480	&	0.00001	&	0.00002	&	0.2	&		&		\\
	&	no peer	&	1485	&	             	&	      	&		&		&		\\
\addlinespace[0.5ex]
10	&	   euclidean 	&	7525	&	   0.00030\sym{***} 	&	0.00006	&	0.1	&	0.095	&	4859	\\
	&	  on-road 	&	7552	&	  0.00022\sym{***} 	&	0.00005	&	0.3	&		&		\\
	&	no peer	&	7579	&	             	&	    	&	      	&		&		\\
\addlinespace[0.5ex]
11	&	  on-road 	&	4050	&	    0.00034\sym{\circ} 	&	0.0002	&	0.1	&	0.137	&	1946	\\
	&	   euclidean 	&	4053	&	    0.00008\sym{\circ} 	&	0.00005	&	0.2	&		&		\\
	&	no peer	&	4071	&	             	&	       	&	       	&		&		\\
\addlinespace[0.5ex]
12	&	   euclidean 	&	4553	&	    0.00025\sym{*} 	&	0.00012	&	0.1	&	0.109	&	2637	\\
	&	  on-road 	&	4554	&	   0.00027\sym{**} 	&	0.0001	&	0.3	&		&		\\
	&	no peer	&	4627	&	             	&	     	&	        	&		&		\\
\addlinespace[0.5ex]
13	&	  on-road 	&	7601	&	  0.00057\sym{***} 	&	0.00011	&	0.2	&	0.148	&	3409	\\
	&	   euclidean 	&	7607	&	  0.00057\sym{***} 	&	0.00012	&	0.1	&		&		\\
	&	no peer	&	7709	&	            	&	      	&	       	&		&		\\
\addlinespace[0.5ex]
14	&	no peer	&	212	&	            	&	       	&		&	0.005	&	1836	\\
	&	   euclidean 	&	217	&	0	&	0.00922	&	0.2	&		&		\\
	&	  on-road 	&	217	&	0	&	0.01138	&	0.3	&		&		\\
\addlinespace[0.5ex]
15	&	   euclidean 	&	13053	&	  0.00023\sym{***} 	&	0.00006	&	0.1	&	0.133	&	6300	\\
	&	  on-road 	&	13062	&	   0.00017\sym{**} 	&	0.00005	&	0.2	&		&		\\
	&	no peer	&	13124	&	             	&	      	&	        	&		&		\\
\addlinespace[0.5ex]
16	&	  on-road 	&	6016	&	  0.00025\sym{***} 	&	0.00008	&	0.2	&	0.051	&	7021	\\
	&	   euclidean 	&	6020	&	   0.00019\sym{**} 	&	0.00006	&	0.1	&		&		\\
	&	no peer	&	6229	&	             	&	       	&		&		&		\\
\bottomrule
\multicolumn{5}{l}{\footnotesize \sym{\circ} \(p<0.1\) \sym{*} \(p<0.05\), \sym{**} \(p<0.01\), \sym{***} \(p<0.001\)}\\
\end{tabular}

%\end{tabular}
\end{table}

\subsection{Prediction Performance}
We measure the predictive performance of the epidemic model using cross validation by dividing the data into a training and test period. We infer the parameters ($\lambda_0^t,\bar{\beta},\alpha$) from the training period and measure the model's performance on the test period. This approach penalizes models with too many parameters: if the model has been overfit in the training period, it will measure a poor performance on the test period. In our case, the model with both endemic and epidemic components has one parameter more than the corresponding model where only an endemic component is allowed. This implies that if the predictive performance measured in the test period is lower for the model with an epidemic effect included, then having one more parameter does not help prediction and we should prefer the model with only an endemic component.
We select Jan 2004 to Dec 2012 as the training period, and the final three years of data (Jan 2013 to Dec 2015) as the test period.
We measure the models predictive performance as the ability to predict how WSL participation evolves through time during the last 36 months and compare with the observed dynamics. We use the parameters fitted on the training set to simulate the future dynamics via Markov chain Monte Carlo simulation. We then calculate the root mean square error between the average participation rate over 100 Monte Carlo realizations $\hat{I}^{sim}=(\hat{I}^{sim}_1,\dots,\hat{I}^{sim}_T)$ and the observed dynamics $I^{obs}=(I^{obs}_1,\dots,I^{obs}_T)$, where $\hat{I}^{sim}_t$ and ${I}^{obs}_t$ are the ratios of participating houses (total number of participating houses divided by the total number of houses in the focal area) at time $t$ from the Monte Carlo simulations and the observed dynamics respectively. Formally:

\begin{equation}\label{eqn:rmse}
RMSE=\sqrt{\frac{1}{T}\sum_{t=1}^T \bup{ \hat{I}^{sim}_t - I^{obs}_t }^2}
\end{equation}
where $t=1,T$ correspond to Jan 2013 and Dec 2015 respectively. We compare simulations results for $\tau_R=12$ and the values of $\tau_d$ on both of on-road and euclidean based networks that have the lowest AIC on the training dataset. As we mentioned before, the model with best AIC might not coincide with the one with best prediction results. However, it is not feasible to implement Monte Carlo simulation for all possible models, thus we make this choice in selecting what model to consider for measuring prediction's performance. 
We then compare performance with the case with no epidemic, i.e. $\alpha=0$.
We did not run Monte Carlo realizations for neighborhood $n=14$ because the inferred epidemic component was null, and so there is no difference between the two models. Notice that the full joint dynamics of the entire system during the test period is not deterministic, therefore we need to consider Monte Carlo simulations starting all from the same initial conditions to estimate its stochastic behavior.
In Tab.~\ref{table:prediction} we can see that for all neighborhoods, the RMSE for the model with an epidemic component is lower than the one for the model with no epidemic, although in some cases the difference is so small that it is insignificant ($n \in \{7,8,9,16 \}$). For these neighborhoods, the final time step activation ratio $\hat{I}^{sim}_T$ for the epidemic model is within a standard deviation from the one with no epidemic, indicating that the two models do not show a significant difference in their predictions of the number of participating houses 36 months after the end of the training set. 

Comparing the temporal dynamics with the results of the inference approach we can argue that for neighborhoods $n\in\{2,3,4,5,10,13,15\}$ the epidemic model is preferred by both approaches, thus providing a strong evidence that the presence of peer effect in the model helps represent WSL adoption dynamics.  For neighborhoods $n\in\{8,9,14 \}$ the evidence is against the presence of peer effect because both methodologies give negative results. For the remaining neighborhoods $n\in\{1,6,7,11,12,16 \}$, the evidence is less clear because one of the two approaches shows a performance improvement for the model including epidemic effects, while the other does not. %However, we should also note that the prediction performance that we obtained could in principle be improved if we would run all possible combinations of $\tau_R$ and $\tau_d$. Some of these could have had higher prediction performance but worse AIC, and therefore we excluded from our tests of prediction. Nevertheless,  it is not feasible to consider all possible models and we believe that the broad conclusions likely would not change: that in some places, . 

The participation ratio predicted at the final time step in Dec 2015 is in general overestimated by both models, indicating the presence of a slow down in adoption dynamics. This could be caused by structural constraints which limit program uptake and are not explicitly observed in our dataset but it is hard to speculate about their precise nature in the absence of further evidence.

\begin{table}[!htbp]\label{table:prediction}
\centering
\caption{Prediction results. Within each neighborhood, results are ordered by the root mean square error (RMSE) (\ref{eqn:rmse}) over 100 Monte Carlo realizations. In all these tests the recovery time is fixed to $\tau_R=12$ months; $\tau_d$ is the distance threshold used to build the network; $I^{obs}_T$ and $\hat{I}^{sim}_T$ are the observed and predicted final-time participation ratios at the end of the study period T. The standard deviation of $\hat{I}^{sim}_T$ is $\sigma_{\hat{I}^{sim}_T}$. }
%\begin{tabular}{lllllllrll}
\begin{tabular}{rlrrrrr}
\toprule
 Neighborhood & Distance Metric &  $\tau_d$ & RMSE &  $\bar{I}^{sim}_T$ &   $\sigma_{\hat{I}^{sim}_T}$ &  $I^{obs}_T$  \\
\midrule
1	&	 euclidean 	&	0.3	&	0.0025	&	0.0536	&	0.0002	&	0.0487	\\
	&	 on-road   	&	0.2	&	0.003	&	0.0544	&	0.0002	&		\\
	&	 no peer   	&		&	0.0032	&	0.0546	&	0.0002	&		\\
\addlinespace[0.8ex]													
2	&	 euclidean 	&	0.3	&	0.0026	&	0.1472	&	0.0003	&	0.1407	\\
	&	 on-road   	&	0.2	&	0.0049	&	0.1514	&	0.0002	&		\\
	&	 no peer   	&		&	0.0055	&	0.1523	&	0.0002	&		\\
\addlinespace[0.8ex]													
3	&	 euclidean 	&	0.2	&	0.0058	&	0.1429	&	0.0003	&	0.1316	\\
	&	 on-road   	&	0.3	&	0.0059	&	0.1429	&	0.0002	&		\\
	&	 no peer   	&		&	0.0074	&	0.1453	&	0.0002	&		\\
\addlinespace[0.8ex]													
4	&	 euclidean 	&	0.2	&	0.0067	&	0.1454	&	0.0002	&	0.1321	\\
	&	 on-road   	&	0.3	&	0.0069	&	0.1455	&	0.0003	&		\\
	&	 no peer   	&		&	0.0077	&	0.147	&	0.0002	&		\\
\addlinespace[0.8ex]													
5	&	 euclidean 	&	0.1	&	0.0045	&	0.1122	&	0.0002	&	0.1035	\\
	&	 on-road   	&	0.1	&	0.0048	&	0.1127	&	0.0002	&		\\
	&	 no peer   	&		&	0.005	&	0.1129	&	0.0002	&		\\
\addlinespace[0.8ex]                  													
6	&	 euclidean 	&	0.3	&	0.0055	&	0.0832	&	0.0002	&	0.0737	\\
	&	 on-road   	&	0.2	&	0.0058	&	0.0836	&	0.0002	&		\\
	&	 no peer   	&		&	0.007	&	0.0856	&	0.0003	&		\\
\addlinespace[0.8ex]          													
7	&	 on-road   	&	0.3	&	0.0034	&	0.0411	&	0.0002	&	0.035	\\
	&	 euclidean 	&	0.1	&	0.0035	&	0.0412	&	0.0002	&		\\
	&	 no peer   	&		&	0.0035	&	0.041	&	0.0002	&		\\
\addlinespace[0.8ex]													
8	&	 on-road   	&	0.1	&	0.0038	&	0.0779	&	0.0005	&	0.0704	\\
	&	 euclidean 	&	0.3	&	0.0039	&	0.0781	&	0.0006	&		\\
	&	 no peer   	&		&	0.0041	&	0.0784	&	0.0005	&		\\
\addlinespace[0.8ex]													
9	&	 on-road   	&	0.3	&	0.001	&	0.0152	&	0.0001	&	0.013	\\
	&	 euclidean 	&	0.2	&	0.0011	&	0.0153	&	0.0001	&		\\
	&	 no peer   	&		&	0.0012	&	0.0154	&	0.0001	&		\\
\addlinespace[0.8ex]            													
10	&	 euclidean 	&	0.1	&	0.0078	&	0.1094	&	0.0002	&	0.0953	\\
	&	 on-road   	&	0.2	&	0.008	&	0.1097	&	0.0002	&		\\
	&	 no peer   	&		&	0.0086	&	0.1106	&	0.0002	&		\\
\addlinespace[0.8ex]           													
11	&	 on-road   	&	0.1	&	0.0058	&	0.1473	&	0.0004	&	0.1367	\\
	&	 euclidean 	&	0.1	&	0.0059	&	0.1477	&	0.0003	&		\\
	&	 no peer   	&		&	0.0069	&	0.1492	&	0.0003	&		\\
\addlinespace[0.8ex]           													
12	&	 euclidean 	&	0.1	&	0.0062	&	0.1217	&	0.0003	&	0.1088	\\
	&	 on-road   	&	0.3	&	0.0064	&	0.122	&	0.0003	&		\\
	&	 no peer   	&		&	0.0068	&	0.1226	&	0.0003	&		\\
\addlinespace[0.8ex]           													
13	&	 on-road   	&	0.3	&	0.0039	&	0.1585	&	0.0003	&	0.1484	\\
	&	 euclidean 	&	0.1	&	0.0041	&	0.1588	&	0.0003	&		\\
	&	 no peer   	&		&	0.005	&	0.1605	&	0.0003	&		\\
\addlinespace[0.8ex]         													
15	&	 euclidean 	&	0.1	&	0.0067	&	0.1468	&	0.0002	&	0.1333	\\
	&	 on-road   	&	0.1	&	0.0072	&	0.1478	&	0.0002	&		\\
	&	 no peer   	&		&	0.0075	&	0.148	&	0.0002	&		\\
\addlinespace[0.8ex] 													
16	&	 euclidean 	&	0.1	&	0.003	&	0.0566	&	0.0001	&	0.051	\\
	&	 on-road   	&	0.2	&	0.0031	&	0.0568	&	0.0001	&		\\
	&	 no peer   	&		&	0.0031	&	0.0567	&	0.0001	&		\\
\bottomrule
\end{tabular}

%\end{tabular}
\end{table}
\section{Conclusions}
In this work we study neighborhood based peer effects around participation in Las Vegas, Nevada's water conservation program, Water Smart Landscapes. 
Using a discrete-time mechanistic epidemic model on different network topologies we find evidence of peer effects in several of the randomly selected neighborhoods considered in this study. 
This indicates that the presence of a participating home increases the likelihood of their neighbors also participating in the program. 
We provided evidence of this effect using both an inference and a prediction approach, and find that for several neighborhoods the epidemic model has both better AIC and better performance in predicting the program's future adoption.
In addition, we find evidence that there is a recovery period in the influence a household's WSL conversion has on neighboring homes.  
Models where an active house impacts its neighbors' decisions to participate for several months to a year are preferred over models where this contagion is indefinite, suggesting that the peer effect fades in time.

We provided a mapping between a mechanistic and discrete-time epidemic model with autoinfection and an additive-multiplicative hazard model; this allows us to capture variations in autoinfection probabilities based on covariates on individual houses.
In both the inference and prediction approaches, we carefully control for the main challenges associated with identifying peer effects:  reflection, homophily, and correlated unobservables.   
The prediction approach is particularly well suited to distinguishing peer effects from homophily and correlated unobservables, because these types of self sorting or environmental factors can be expected to change slowly relative to changes in WSL participation: moving between homes is a significantly larger decision than changing the outdoor landscape in that home.
Thus, this type of epidemic modeling can extend and complement hazards models that have been used in economics to identify the existence of peer effects. 

Applying techniques from epidemic modeling to the analysis of peer effects suggests opportunities to explore a wide range of  problems related to the one studied here by exploiting recent methodological developments in inference and optimization \cite{altarelli2014bayesian,altarelli2014containing}. 
For instance, these methods could be used to seek an optimal set of houses to target through a marketing campaign in order to maximize the spread of WSL adoption. 
Characteristics of neighborhoods where evidence suggests epidemic spreading is not occurring could also be studied in order to explore any structural impediments to participation, information water agencies may find useful for program planning. 
In this work we focused on a direct measure of contagion effect by considering program adoption as main observable quantity. 
It would be interesting to extend these findings to indirect measures, such as effects of one households WSL participation on their neighbors water consumption.

\section{Acknowledgements}

We thank Alfredo Braunstein and Joshua K Abbott for helpful comments. Research sponsored by the Laboratory Directed Research and Development Program of Oak Ridge National Laboratory, managed by UT-Battelle, LLC, for the US Department of Energy. CB also received partial support from the ASU/SFI Center for Biosocial Complexity.  CDB was supported by the John Templeton Foundation.

\section{Appendix}
\subsection{Log Likelihood of the Epidemic Model}\label{apx:logL}
\bea \label{eq:logL}
\mathcal{L}(\bar{t}|\alpha,\mu)&=&\sum_{i \in V | x_i^T=S} \rup{ \sum_{t=1}^{T} \log (1-\mu_i^t) + \sum_{k \in \partial i | t_{kI}<T} ({T-t_{kiI}}) \log (1-\alpha)} +\\
&&+\sum_{i \in V | x_i^T\neq S} \left[  \sum_{k \in \partial i | t_{kI}<t_{iE}-1} ({t_{iE}-\tau_{kiI}-1}) \log (1-\alpha) +\log \rup{1-(1-\mu_i^{t_{iE}})(1-\alpha)^{n_{i}}} \right. \nonumber \\
&& \left.+\sum_{t=1}^{t_{iE}-1} \log (1-\mu_i^t) \right]  \nonumber
\eea
\subsection{Mapping Epidemic with Autoinfection to Additive-Multiplicative Hazard Model}\label{apx:epidemichazard}
Here we outline how the three models SI, additive-multiplicative and multiplicative hazard, describe the same quantity of interest, the hazard rate $\lambda (t_{i} | \Theta, D)$. In discrete time, this is the \textit{conditional} probability that an event happens at time $t_{i}$, \textit{given} it has not yet happened before. In general this quantity can depend on a set of parameters $\Theta$ and the data $D$. In our case $\Theta=\{ \alpha, \mu\}$ include the transmission and autoinfection probabilities of all nodes, whereas $D=\{ t_{iE}\}$, i.e. the data are the observed exposure times.\\
In the SEIR model, ($1$ minus) the hazard rate of a susceptible node $i$ at time $t$ is:
\be\label{lambdaSI}
1-\lambda_i (t | \Theta, D)= (1-\alpha)^{n_{i}^{t}} \,(1-\mu_{i}^{t})    \qquad \mbox{Epidemic model}
\ee

In discrete time \cite{kalbfleisch2011statistical}, the hazard rate
in the additive-multiplicative hazard model \cite{hohle2005inference,hohle2008spatio} can be written as:
\be\label{additiveHohle}
1-\lambda_i(t|\Theta,D) = \bup{1-\alpha}^{n_{i}^{t}} \, \bup{e^{-\lambda_{0}^t}}^{e^{ \bar{x}_{i} \cdot \bar{\beta}}}  \qquad \mbox{Additive-Multiplicative}
\ee
whereas, in the purely multiplicative model used in \cite{towe_contagion_2013} we have:
\be \label{mutliplicativeTowe}
1-\lambda_{i}(t|\Theta,D) =\bup{e^{-\lambda_{0}^t}}^{e^{ \bar{x}_{i}  \cdot \bar{\beta}} e^{ \alpha \, n_{i}^{t}} }    		\:\:\:\:\: \qquad \qquad \mbox{Multiplicative}
\ee
where $\lambda_0^t$ is the baseline hazard at time $t$ which represents a global contribution to the probability of getting infected which is the same for all houses; $\bar{x}_i$ is a vector of covariates and $\bar{\beta}$ is a vector of parameters coupling the covariates, in a similar flavor as in linear regression.\\
Comparing (\ref{lambdaSI}) and (\ref{additiveHohle}) we obtain the mapping (\ref{eq:epihaz}).

\subsubsection*{Proof}
From the definition of hazard rate \cite{kalbfleisch2011statistical} which, in discrete time, is a conditional probability, we have a relationship connecting the probability $f(t_{i})$ of an event happening at time $t_{i}$ and the survival probability $S(t_{i})$, which is the probability that no event happens before time $t_{i}$ (but it can happen at exactly $t_{i}$ or later):

\be\label{lambdaFS}
\lambda(t)=\f{f(t)}{S(t)} 
\ee

Notice that for discrete time $S(t)=\sum_{s=t}^{\infty} f(s)$,  thus we can write:
\be
f(t)= S(t)-S(t+1) =\sum_{s=t}^{\infty} f(s)-\sum_{s=t+1}^{\infty} f(s)= f(t)
\ee
 Substituting into (\ref{lambdaFS}):
 \bea
\lambda(t)&=&\f{f(t)}{S(t)} \\
&=& 1-\f{S(t+1)}{S(t)} \label{lambdaS}
\eea

Using the equation valid in general \cite{kalbfleisch2011statistical} relating the survival probability $S(t)$ with the hazard rate $S_i(t)=e^{-\int_{0}^{t} \lambda_i(s) \, ds}$, the following relation valid for additive-multiplicative hazard model as in \cite{scheike2002additive}, where $\lambda_i^t=\alpha n_i^t+\lambda_0 e^{\bar{x_i} \cdot \bar{\beta}}$ with covariates can be derived:
\be \label{survivalfunction}
S_i(t)=S_0(t)^{\exp(\bar{x_i} \cdot \bar{\beta})} \,\, S^i_{epi}(t)
\ee
where $S_0(t)$ represents the baseline survival probability when $\bar{\beta}=0$ and no epidemic effect is present ($\alpha=0$), whereas $S^i_{epi}(t)$ represents the epidemic contribution to the survival probability.\\
For discrete models we can write:
\begin{eqnarray}
S_0(t)&=&\prod_{s=0}^{t-1} (e^{-\lambda_0^s}) \label{survival0}\\
S^i_{epi}(t) &=&\prod_{s=0}^{t-1} (1-\alpha)^{ n_i^s} \label{survivalepi}
\end{eqnarray}

Substituting (\ref{survival0}) and (\ref{survivalepi}) into (\ref{lambdaS}), we get:
 \bea
\lambda_i^t&=&1-(e^{-\lambda_0^s})^{\exp(\bar{x_i} \cdot \bar{\beta})} \, (1-\alpha)^{ n_i^t} 
\eea

For the purely Multiplicative hazard model as in \cite{towe_contagion_2013} where $\lambda_i^t=\lambda_0^t \, e^{\bar{x_i} \cdot \bar{\beta}} \,e^{\alpha n_i^t}$ we cannot separate the baseline and epidemic contribution as above, instead we have:
\begin{equation}
S_i(t)=S_0(t)^{\exp(\bar{x_i} \cdot \bar{\beta} +\alpha n_i^t)} 
\end{equation}
 which leads to:
\bea
\lambda_i^t&=&1-(e^{-\lambda_0^s})^{\exp(\bar{x_i} \cdot \bar{\beta}+\alpha n_i^t)}  
\eea

\begin{table}[!htbp]\label{table:inference_recovery}
\begin{adjustwidth}{-0.9in}{-0.9in}% adjust the L and R margins by 1 inch
\def\sym#1{\ifmmode^{#1}\else\(^{#1}\)\fi}

\centering
\caption{Maximum likelihood inference results for different recovery times $\tau_R$ (in months). $\tau_d$ is the distance threshold used to build the network, $n$ is the neighborhood's id; the subscript under the $AIC$ refers to the corresponding value of $\tau_R$; the third line of every neighborhood refers to the model with no epidemic ($\tau_d=0.0$ and $\alpha=0$). }
%\begin{tabular}{lllllllrll}
\begin{tabular}{llllllllllll}
\toprule
n &          $\tau_R=1$ &          $\tau_R=4$  &          $\tau_R=6$  &         $\tau_R=12$  &      $\tau_R=\infty$  &   $AIC_1$ &    $AIC_4$ &    $AIC_6$ &   $AIC_{12}$ &  $AIC_\infty$ &        dist \\
\midrule
          1 &    0.00071 &         0.0 &     0.00011 &    0.00008\sym{*} &     7e-06 &   2860 &   2867 &   2863 &   2855 &    2880 &   euclidean \\
          1 &    0.00024 &      0.0002 &         0.0 &         0.0 &         0.0 &   2830 &   2866 &   2867 &   2867 &    2867 &  on-road \\
          1 &        0.0 &         0.0 &         0.0 &         0.0 &         0.0 &   2865 &   2865 &   2865 &   2865 &    2865 &    \\
 \addlinespace[0.8ex]         
          2 &   0.00023\sym{*} &   0.00020\sym{***} &  0.00018\sym{***} &  0.00012\sym{***} &     5e-06 &  10829 &  10807 &  10792 &  10786 &   10833 &   euclidean \\
          2 &    0.00039 &   0.0003\sym{***} &  0.00023\sym{***} &   0.00020\sym{***} &     0.00002 &  10834 &  10823 &  10823 &  10813 &   10834 &  on-road \\
          2 &        0.0 &         0.0 &         0.0 &         0.0 &         0.0 &  10835 &  10835 &  10835 &  10835 &   10835 &    \\
\addlinespace[0.8ex]
          3 &   0.00015\sym{\circ} &  0.00026\sym{***} &  0.00013\sym{***} &  0.00012\sym{***} &         0.0 &  10994 &  10976 &  10972 &  10977 &   11042 &   euclidean \\
          3 &   0.00039\sym{*} &  0.00035\sym{***} &  0.00032\sym{***} &  0.00021\sym{***} &     9e-06 &  10989 &  10977 &  10970 &  10967 &   11042 &  on-road \\
          3 &        0.0 &         0.0 &         0.0 &         0.0 &         0.0 &  11040 &  11040 &  11040 &  11040 &   11040 &  \\
\addlinespace[0.8ex]          
          4 &     0.00010 &  0.00023\sym{***} &   0.00020\sym{***} &  0.00010\sym{***} &         0.0 &  12431 &  12411 &  12405 &  12415 &   12473 &   euclidean \\
          4 &   0.00061\sym{*} &  0.00033\sym{***} &  0.00032\sym{***} &  0.00025\sym{***} &         0.0 &  12426 &  12411 &  12400 &  12412 &   12473 &  on-road \\
          4 &        0.0 &         0.0 &         0.0 &         0.0 &         0.0 &  12471 &  12471 &  12471 &  12471 &   12471 &   \\
\addlinespace[0.8ex]          
          5 &    0.00015 &   0.00014\sym{**} &    0.00022\sym{*} &   0.00017\sym{**} &         0.0 &   8408 &   8389 &   8397 &   8402 &    8428 &   euclidean \\
          5 &     0.00001 &    0.00024\sym{\circ} &    0.00018\sym{\circ} &   0.00015\sym{**} &         0.0 &   8401 &   8407 &   8407 &   8415 &    8428 &  on-road \\
          5 &        0.0 &         0.0 &         0.0 &         0.0 &         0.0 &   8426  &   8426  &   8426  &   8426  &    8426 &    \\           
 \addlinespace[0.8ex]         
          6 &    0.00048 &    0.00037\sym{*} &    0.00022\sym{\circ} &    0.00006\sym{*} &     0.00003 &   5089 &   5087 &   5087 &   5087 &    5091 &   euclidean \\
          6 &    0.00077 &    0.00036\sym{*} &     0.00004 &         0.0 &     0.00003 &   5093 &   5084 &   5092 &   5095 &    5091 &  on-road \\
          6 &        0.0 &         0.0 &         0.0 &         0.0 &         0.0 &   5093 &   5093 &   5093 &   5093 &    5093 &    \\
\addlinespace[0.8ex]          
          7 &     0.00020 &  0.00043\sym{***} &  0.00019\sym{***} &  0.00026\sym{***} &    0.00007\sym{***} &   2253 &   2236 &   2238 &   2221 &    2255 &   euclidean \\
          7 &    0.00035 &    0.00055\sym{*} &    0.00029\sym{*} &  0.00027\sym{***} &   0.00012\sym{**} &   2259 &   2236 &   2251 &   2211 &    2218 &  on-road \\
          7 &        0.0 &         0.0 &         0.0 &         0.0 &         0.0 &   2328 &   2328 &   2328 &   2328 &    2328 &   \\
 \addlinespace[0.8ex]         
          8 &      0.00001 &     0.00037 &     0.00035 &     0.00016 &         0.0 &    835 &    834 &    835 &    835 &     838 &   euclidean \\
          8 &      0.00001 &     0.00051 &     0.00028 &     0.00011 &         0.0 &    835 &    833 &    834 &    835 &     838 &  on-road \\
          8 &        0.0 &         0.0 &         0.0 &         0.0 &         0.0 &    836 &    836 &    836 &    836 &     836 &    \\          
 \addlinespace[0.8ex]         
          9 &        0.0 &     0.00018 &     0.00021 &       6e-06 &     2e-06 &   1487 &   1471 &   1474 &   1480 &    1495 &   euclidean \\
          9 &        0.0 &     0.00023 &    0.00067\sym{\circ} &    0.00012\sym{\circ} &     0.00008 &   1487 &   1460 &   1470 &   1463 &    1460 &  on-road \\
          9 &        0.0 &         0.0 &         0.0 &         0.0 &         0.0 &   1485 &   1485 &   1485 &   1485 &    1485 &   \\ 
 \addlinespace[0.8ex]        
         10 &    0.00011 &  0.00044\sym{***} &  0.00035\sym{***} &   0.0003\sym{***} &         0.0 &   7554 &   7538 &   7537 &   7525 &    7581 &   euclidean \\
         10 &   0.00077\sym{*} &   0.00037\sym{**} &    0.00036\sym{*} &  0.00022\sym{***} &         0.0 &   7548 &   7546 &   7551 &   7552 &    7581 &  on-road \\
         10 &        0.0 &         0.0 &         0.0 &         0.0 &         0.0 &   7579 &   7579 &   7579 &   7579 &    7579 &    \\
\addlinespace[0.8ex]         
         11 &    0.00013 &     0.00015 &    0.00014\sym{\circ} &    0.00008\sym{\circ} &         0.0 &   4053 &   4054 &   4052 &   4053 &    4073 &   euclidean \\
         11 &    0.00026 &      0.00010 &     0.00041 &    0.00034\sym{\circ} &     0.00001 &   4053 &   4053 &   4051 &   4050 &    4072 &  on-road \\
         11 &        0.0 &         0.0 &         0.0 &         0.0 &         0.0 &   4071 &   4071 &   4071 &   4071 &    4071 &   \\
\addlinespace[0.8ex]         
         12 &      1e-05 &      0.0001 &     0.00023 &    0.00025\sym{*} &     0.00005 &   4565 &   4568 &   4572 &   4553 &    4566 &   euclidean \\
         12 &    0.00024 &     0.00019 &    0.00028\sym{*} &   0.00027\sym{**} &     7e-06 &   4565 &   4564 &   4564 &   4554 &    4579 &  on-road \\
         12 &        0.0 &         0.0 &         0.0 &         0.0 &         0.0 &   4627 &   4627 &   4627 &   4627 &    4627 &    \\
\addlinespace[0.8ex]         
         13 &   0.00160\sym{**} &  0.00027\sym{***} &  0.00024\sym{***} &  0.00057\sym{***} &   0.00009\sym{**} &   7620 &   7615 &   7604 &   7607 &    7663 &   euclidean \\
         13 &  0.00091\sym{**} &   0.00044\sym{**} &  0.00065\sym{***} &  0.00057\sym{***} &   0.00019\sym{**} &   7627 &   7619 &   7613 &   7601 &    7614 &  on-road \\
         13 &        0.0 &         0.0 &         0.0 &         0.0 &         0.0 &   7709 &   7709 &   7709 &   7709 &    7709 &   \\
\addlinespace[0.8ex]         
         15 &   0.00025\sym{*} &  0.00044\sym{***} &  0.00034\sym{***} &  0.00023\sym{***} &         0.0 &  13067 &  13053 &  13054 &  13053 &   13126 &   euclidean \\
         15 &    0.00016 &    0.00049\sym{*} &   0.00025\sym{**} &   0.00017\sym{**} &     0.00002 &  13073 &  13065 &  13062 &  13062 &   13120 &  on-road \\
         15 &        0.0 &         0.0 &         0.0 &         0.0 &         0.0 &  13124 &  13124 &  13124 &  13124 &   13124 &    \\
\addlinespace[0.8ex]        
         16 &    0.00007 &     0.00016 &    0.00022\sym{*} &   0.00019\sym{**} &  0.00004\sym{***} &   6031 &   6037 &   6024 &   6020 &    6140 &   euclidean \\
         16 &    0.00011 &     0.00012 &   0.00032\sym{**} &  0.00025\sym{***} &     0.00005 &   6030 &   6072 &   6017 &   6016 &    6031 &  on-road \\
         16 &        0.0 &         0.0 &         0.0 &         0.0 &         0.0 &   6105 &   6105 &   6105 &   6105 &    6105 &   \\
\bottomrule
\multicolumn{5}{l}{\footnotesize \sym{\circ} \(p<0.1\) \sym{*} \(p<0.05\), \sym{**} \(p<0.01\), \sym{***} \(p<0.001\)}\\
\end{tabular}
%\end{tabular}
\end{adjustwidth}

\end{table}

%\subsection{Neighborhood logistic}\label{App: neighborhood logistic}
%\input{logistic_seed1_7}
%\input{logistic_seed8_12}
%\input{logistic_seed13_20}

%\begin{figure}[h]
%\centering
%\includegraphics[width=12cm]{WSL_area}
%\caption{Average WSL conversion size, separated by the year in which the WSL conversion occurred (x-axis) and homes constructed before (grey) or after (purple) the 2003 restrictions on turf in new construction were in force.}
%\label{fig: WSL conversion size}
%\end{figure}
\clearpage

\bibliographystyle{apalike}
%\bibliography{sample}
\bibliography{peereffects}

\end{document}